\documentstyle[preprint,aps,eqsecnum]{revtex}

\tighten

\begin{document}

\title{Perturbation schemes for systems of nucleons and pions: The
relationship of covariant perturbation theory, the convolution
integral and time-ordered perturbation theory.}

\author{D.~R.~Phillips and I.~R.~Afnan} \address{School of Physical
Sciences, The Flinders University of South Australia,\\ GPO Box 2100,
Adelaide 5001, Australia.}

\date{\today}

\maketitle

\begin{abstract} This paper is the first in a series of three which
attempt to resolve the difficulties that have plagued the $NN-\pi NN$
problem for the past ten years. The problems may be summarized by
saying that the current $NN-\pi NN$ equations {\em cannot} fit the
experimental data. Various theoretical inconsistencies in the current
formulation have been pointed out and this work aims to eliminate
these inconsistencies and so, we hope, produce agreement with
experiment. This is to be done by using covariant perturbation
theory, in which these inconsistencies are not present. The covariant
perturbation theory is developed starting from a model Lagrangian, in
order to fix notation and phases. (The approach is actually
Lagrangian independent but we use a specific Lagrangian for
concreteness.) It is shown that both old-fashioned "time-ordered"
perturbation theory and the convolution integral of Kvinikhidze and
Blankleider may be recovered from the covariant perturbation theory
when certain approximations are made. The connection of these results
with the work of Klein, L\'evy, Macke and Kadyshevsky is discussed.
Two forthcoming papers will pursue this covariant calculation in the
$NN-\pi NN$ system using the model and perturbation scheme developed
in this paper and derive fully covariant $NN-\pi NN$ equations
without the double counting problems present in previous covariant
equations.        \end{abstract}

\pacs {PACS number(s):11.10.Ef, 13.75.Cs, 13.75.Gx, 21.30.+y,
21.40.+d, 24.10.Cn}

\bibliographystyle{PhysicalReview}

\section {Introduction}

\label {sec-intro}

Models in nuclear and particle physics are of two main types:
(i)~Those based directly on Quantum Chromodynamics (QCD), in which
quarks are the basic degrees of freedom, e.g. Chiral Perturbation
Theory \cite{GL82,Ga90} or the Nambu-Jona-Lasinio model
\cite{NJL61A,NJL61B,Kl92,We93}; and (ii)~Those in which mesons and
baryons are the basic degrees of freedom, e.g. the Walecka models
of the nuclear many-body system, which are collectively known as
Quantum Hadrodynamics~(QHD) \cite{Wa74,Se92}.  In principle these
two types of models are connected, since the Lagrangian for QHD
should be obtained from that of QCD upon integration of the
quark-gluon degrees of freedom. However, as yet this has not been
achieved, and one has to resort to models of QCD such as chiral bag
models \cite{Th84} in order to explain the connection between the
quark and the hadronic degrees of freedom. Since the hadronic
picture can in this way be justified from the quark picture one
would expect that the hadronic models would be adequate
approximations to the true physical situation. Whether this is the
case or not can only be determined by comparison of the predictions
of the hadronic models with experiment. A disagreement between
theory and experiment can then be due either to some approximation
in the calculation using the model, or to the need for explicit
quark-gluon degrees of freedom in the model. With the current state
of computer technology, light nuclear systems are the only nuclear
systems in which we may perform the calculation of experimentally
measured quantities while retaining some control over the
approximations used. Consequently, only in light nuclear systems
may we discover whether the models of the second type, in which
mesons and baryons are the basic degrees of freedom, can accurately
describe the properties of nuclei. If they cannot, it will be
necessary to resort to models of the first type, with explicit
quark degrees of freedom, in order to obtain an accurate
description of nuclear properties. Therefore, the interaction of
pions and photons with light nuclear systems is a testing ground
for models in particle and nuclear physics.

The $NN-\pi NN$ system is one example of a light nuclear system in
which this program for the testing of models based on meson-baryon
degrees of freedom may be pursued. During the last twenty years
disagreements between theory and experiment have led to an
improvement in the calculations based on meson-baryon degrees of
freedom. The physics content of these models has gradually
increased as they attempt to reproduce the experimental data ever
more closely. To illustrate how experimental results and the need
for consistency in the theory have together influenced the
development of these $NN-\pi NN$ models, we will briefly review the
history of the $NN-\pi NN$ equations. This will also allow us to
focus on some of the problems which led to the current
investigation. A detailed analysis of the theoretical methods and a
comparison of their predictions with experiment may be found in the
recent book by Garcilazo and Mizutani~\cite{GM90}.

The suggestion that the one pion exchange potential could be
treated within the framework of the Faddeev equations was first
made by Varma who calculated $NN$ scattering using a one pion
exchange potential and the Faddeev equations~\cite{Va67}. In this
model, the absorption and production of the pion is the result of
the $\pi N$ amplitude in the $P_{11}$ channel having a bound state
pole with binding energy equal to the pion mass. This model was
later extended by Afnan and Thomas to include pion production in
$NN$ scattering, pion absorption in $\pi d$ scattering and $\pi d$
elastic scattering~\cite{AT74}. In this way a unified formulation
of the three reactions:  \begin{eqnarray} N + N &\rightarrow& N + N
\nonumber, \\* N + N &\leftrightarrow& \pi +
d,                        \label{eq:1.1}\\* \pi + d &\rightarrow&
\pi + d \nonumber   \end{eqnarray} was achieved for the first time.
The problem with this model was that the nucleons were not
identical: one of the nucleons (labeled the $N'$) was treated as a
$\pi N$ bound state and so could emit a pion, while the other was
an elementary nucleon, and so could {\em not} emit a pion.

However, the success of the model in predicting the $NN$ phase
shifts for large angular momenta, and the ability of the model to
describe $\pi d$ scattering by not only summing the multiple
scattering series, but also including the effects of real
absorption, motivated several groups to reformulate the model in
such a way that both nucleons could emit a pion.  All of these
formulations were based on a field theory allowing the possibility
of an arbitrary number of pion emissions and absorptions. However,
the constraint of unitarity allowed all the groups to truncate the
field theory. Some groups performed this truncation using Feshbach
projection operators~\cite{Fe58,Fe62}, while others did it through
the classification of diagrams technique of Taylor~\cite{Ta63}. All
groups truncated the theory by neglecting or approximating
processes involving four or more particles---the so-called one-pion
approximation---and so obtained a set of linear integral equations,
known as the $NN-\pi NN$ equations, which are three-dimensional in
momentum
space~\cite{Th73,Mi76,MK77,Ri77,TR79,AM79,AM80,AM81,AM83,AB81}. The
equations were then solved and a detailed comparison made with the
extensive experimental data generated in the 1980s. Early attempts
at extending the Hilbert space to include the $\pi\pi NN$ states,
i.e. truncating the field theory at the next stage, led to a more
complicated set of equations which included contributions from the
diagram in Fig.~\ref{fig-Jennings}~\cite{SS78A,SS78B,SS78C}. These
equations simplify to the $NN-\pi NN$ equations when the diagram in
Fig.~\ref{fig-Jennings} is excluded~\cite{AS81}.

Although these Faddeev-type $NN-\pi NN$ equations resolved the
problem of non-identical nucleons by dressing both fermions, the
dressing of the two-nucleon propagator was still not complete. The
truncation of the Hilbert space and the use of time-ordered
perturbation theory meant that both nucleons could not be dressed
at the same time. I.e., in the one-pion approximation, the diagram
on the right of Fig.~\ref{fig-NNdress} was included as part of the
dressing for the $NN$ propagator, but the diagram on the left of
the figure was not---in spite of the fact that the two diagrams are
of the same order in the coupling constant, and are just different
relative time-orders of the same process. This incompleteness of
the dressing, first pointed out by Sauer~et~al.~\cite{Sa85}, gives
rise to a weakening of the $\pi NN$ coupling constant as the number
of nucleons in the system increases. In fact, Sauer~et~al. showed
that in models which, like the $NN-\pi NN$ equations, use an
incomplete dressing of the multi-nucleon propagator, the omission
of some of the dressing means that the $\pi NN$ coupling constant
goes to zero in the nuclear matter limit. Furthermore, the omission
of this dressing also leads to a severe underestimation of the $pp
\leftrightarrow \pi d$ cross-section in theoretical
calculations---especially in theories which treat the $\pi N$
amplitude in the $P_{11}$ channel properly, i.e. as the sum of a
pole and non-pole term. The only way this underestimation can be
remedied is to dress both nucleons fully. Recently, Kvinikhidze and
Blankleider have addressed the question of how to do this and have
elegantly demonstrated how the complete dressing can be achieved
with the use of a convolution integral representation for the $NN$
propagator~\cite{BK92A}.

A second associated problem has plagued all calculations based on
the $NN-\pi NN$ equations which use a $P_{11}$ $\pi N$ amplitude
that is the sum of a pole and non-pole contribution. None of these
calculations has been able to accurately predict the tensor
polarization, $T_{20}$, for $\pi-d$ scattering. In 1988 Jennings
suggested that this discrepancy between theory and experiment is a
result of the $NN-\pi NN$ equations not including the diagram in
Fig.~\ref{fig-Jennings}, which has since become known as the
Jennings mechanism~\cite{Je88,JR88}. Indeed, the mechanism depicted
in Fig.~\ref{fig-Jennings} is merely a different time-order of a
diagram which is included in the $NN-\pi NN$ equations, shown in
Fig.~\ref{fig-Included}. Jennings pointed out that the excluded
diagram may well cancel part of the effect of the one included in
the $NN-\pi NN$ equations, thus producing the correct result for
the tensor polarization in $\pi-d$ scattering.

Since the Jennings mechanism and the extra dressing of the $NN$
propagator are merely different time-orders of diagrams already
included in the $NN-\pi NN$ equations, logically they too should be
included in any calculation of the $NN-\pi NN$ system. Otherwise,
the processes which are occurring on one nucleon restrict the
processes which may occur on the other nucleon. Further, the
failure of the $NN-\pi NN$ equations to correctly describe the
experimental data suggests not only that these two omitted
mechanisms should now be included, but that they {\em must} now be
included.

The omission of these two diagrams from the present theory is a
result of using a truncation based on time-ordered perturbation
theory and unitarity. One might think that the problem could be
resolved by truncating the Hilbert Space at two-pion states, since
then the two diagrams discussed above would both be included. As
mentioned above, an example of a theory which takes this approach
is the model of Stingl and Stelbovics~\cite{SS78A,SS78B,SS78C}.
However, the equations obtained from this model are computationally
difficult to solve. The model also suffers from similar problems to
those of the current $NN-\pi NN$ equations, but at the three-pion
level. One example of this is that diagrams such as those on the
left of Figure~\ref{fig-SSprob} are included in the model, whereas
diagrams such as those on the right are not included, because they
involve a three-pion intermediate state. Again, this occurs even
though the two diagrams are just different time-orders of the same
process.

Clearly then the problem rests not so much with the choice of
truncation point as with a failure to sum over all time-orders when
doing the truncation. Therefore, if we wish to include all the
relevant physics in our model we must use a  perturbation scheme in
which all time-orders are included automatically.  This paper
presents such a perturbation scheme as follows.

In Section~\ref{sec-Lagrangian} the field operators we will be
using are introduced and the Lagrangian for the theory stated. In
Section~\ref{sec-SingleptclGF} we explain how to calculate the
single-nucleon and single-pion Green's functions with this
Lagrangian. At this point we introduce an approximation common in
intermediate-energy nuclear physics---we ignore anti-nucleonic
degrees of freedom. Once we make this approximation, the theory
simplifies considerably. We will show that, apart from this
approximation, the perturbation scheme developed is a fully
covariant theory of nucleons and pions as given in Bjorken and
Drell~\cite{BD64}. In Section \ref{sec-TwonucleonGF} we turn our
attention to the two-nucleon Green's function. It is explained how
to evaluate $G_{NN} (p_1',p_2';p_1,p_2)$, and the Fourier Transform
of the equal-time Green's function $\overline{G}_{NN} (t',
\vec{x}_1^{\, \prime}, \vec{x}_2^{\, \prime}; t, \vec{x}_1,
\vec{x}_2)$ is calculated. We find  that:   \begin{equation}
\overline{G}_{NN} (E', \vec{p}_1^{\, \prime}, \vec{p}_2^{\,
\prime};E,\vec{p}_1, \vec{p}_2)=\int \frac {dz dz'}{(2 \pi)^2}
G_{NN} (E'-z', \vec{p}_1^{\, \prime}, z', \vec{p}_2^{\, \prime};
E-z, \vec{p}_1, z, \vec{p}_2),  \label{eq:CI}  \end{equation}
where $G_{NN} (E'-z',\vec{p}_1^{\, \prime},z',\vec{p}_2^{\,
\prime};E-z,\vec{p}_1,z,\vec{p}_2)$ is calculated using the Feynman
Rules for  $G_{NN} (p_1',p_2';p_1,p_2)$. In Section
\ref{sec-mm'ptclGF} it is shown that similar results hold for the
$m \rightarrow m'$ particle Green's function, a result of
particular relevance to the $\pi NN$ system. This use of a
convolution integral to calculate  multi-particle Green's functions
bears a strong resemblance to the work of Kvinikhidze and
Blankleider (KB)~\cite{BK92A,BK92B} and in Section
\ref{sec-BKconnection} we explain how to derive KB's result from
our field theory. Since KB's work uses a convolution integral in
order to sum a set of time-ordered (or old-fashioned) perturbation
theory diagrams it is necessary to first derive time-ordered
perturbation theory from our covariant perturbation scheme. KB's
result then becomes a consequence of Equation~(\ref{eq:CI}) and the
relationship of time-ordered perturbation theory to the covariant
perturbation scheme. Once this result is established, the
approximations used by KB in order to derive equations for the
$NN-\pi NN$ system~\cite{BK92B} are critically examined, and are
found to have considerable inconsistencies. In Section
\ref{sec-LSZredn} we consider how to calculate amplitudes from the
Green's function $G_{NN}(E'-z',\vec{p}_1^{\,
\prime},z',\vec{p}_2^{\, \prime};E-z,\vec{p}_1,z,\vec{p}_2)$  using
Lehmann-Symanzik-Zimmermann (LSZ) reduction \cite{LSZ55}. We find
that the amplitude generated by a sum of all relative time-orders
of time-ordered perturbation theory diagrams may be calculated by
evaluating the amplitude given by the Feynman rules in
Sections~\ref{sec-Lagrangian} and~\ref{sec-SingleptclGF}. However,
as mentioned above, the perturbation scheme that generates these
Feynman rules is merely an equal-time, no-anti-nucleon
approximation to the full covariant perturbation theory. Therefore,
if we are to include the missing diagrams in our theory we need to
use some sort of covariant perturbation theory in order to derive
new four-dimensional $NN-\pi NN$ equations, even if at certain
stages in the calculation we make approximations.

The results described in this article are similar to those obtained
by Klein, L\'evy, Macke and Kadyshevsky (KLMK)
\cite{Sa52,Le52A,Le52B,Kl53,Kl54,Kl58,KM58,Ma53A,Ma53B,KL74,Ka68,It70}.
The essential difference between KLMK's work and ours is that the
KLMK method is driven by the desire to obtain a three-dimensional
integral equation, whereas one of the main theses presented in this
article is that four-dimensionality is an essential feature of any
theory which is not going to treat different time-orders of the
same physical process differently. We conclude our arguments for
this thesis in Section \ref{sec-KLMK} by briefly describing the
work of KLMK in the context of the calculations performed in the
previous sections. In particular, we highlight the pieces of
physics which are missing from their work, due to their use of a
three-dimensional equation rather than a four-dimensional one.

Haberzettl has also recently published some work on the the use of a
convolution integral in $N$-body scattering, and has derived
four-dimensional integral equations for certain scattering problems
\cite{Ha92,Ha93A,Ha93B}. While his work appears similar to that
discussed here it is approached from a rather different
perspective, being based on cluster dynamics rather than field
theory.

In summary then, the inability of the current $NN-\pi NN$ equations
to correctly describe the experimental data forces the introduction
of new physics into the theory of the $NN-\pi NN$ system: namely,
the consideration of different time-orders of diagrams already
included in the theory. The only natural way to include this
missing physics is to use the full covariant perturbation theory
developed in this paper. The use of such a perturbation theory
will, of course, result in four-dimensional integral equations. Any
attempt to reduce the dimensionality of these equations results in
a loss of physics, as in the work of KB and KLMK. Therefore, in
later papers we pursue the use of the full covariant perturbation
theory and so derive covariant equations for the $NN-\pi NN$
system. In this paper we wish only to concentrate on the
perturbation scheme we shall be using, and to establish the
approximations in which it reduces to the convolution integral
approach, the time-ordered perturbation theory and the work of KB
and KLMK.

\section {The Lagrangian}

\label{sec-Lagrangian}

In this section we define the field theory we shall be using in order
to fix the notation and normalization used throughout the paper. We
define nucleon and meson creation and annihilation operators, state
the commutation relations they obey and write down the Lagrangian and
hence the Hamiltonian. The Lagrangian we use is similar to that used
in Quantum Hadrodynamics calculations \cite{Wa74,Se92}. Note that
although we use a specific Hamiltonian the approach pursued in this
paper could equally well be applied to any meson-baryon Hamiltonian
with vertices which allow the baryons to emit mesons.

Consider second-quantized nucleon field operators $\psi(\vec{x})$ and
$\overline{\psi}(\vec{x})$. We can, of course, Fourier decompose
these Schr\"odinger representation operators, as done by, for
example, Itzykson and Zuber \cite{IZ84}, and when we do we obtain:
\begin {eqnarray} \psi(\vec{x})=\int \frac {d^3q}{(2 \pi)^3}
\frac{m}{E_N(\vec{q})} \sum_{\alpha} [b_\alpha(\vec{q}^{\,})
u_\alpha(\vec{q}^{\,}) e^{i \vec{q} \cdot \vec{x}} +
d^{\dagger}_\alpha(\vec{q}^{\,}) v_\alpha(\vec{q}^{\,}) e^{-i \vec{q}
\cdot \vec{x}}], \label{eq:psi}\\  \overline{\psi}(\vec{x})=\int
\frac {d^3q}{(2 \pi)^3} \frac{m}{E_N(\vec{q}^{\,})} \sum_{\alpha}
[b^{\dagger}_\alpha(\vec{q}^{\,}) \overline{u}_\alpha(\vec{q}^{\,})
e^{-i \vec{q} \cdot \vec{x}} + d_\alpha(\vec{q}^{\,})
\overline{v}_\alpha(\vec{q}^{\,}) e^{i \vec{q} \cdot \vec{x}}] ,
\label {eq:psib} \end {eqnarray} where $m$ is the nucleon mass;
$E_N(\vec{q}^{\,})$ is the energy of a nucleon of momentum $\vec{q}$
and $\alpha$ is a collective label for spin and isospin. Here
$b^{\dagger}(\vec{q}^{\,})$ and $b(\vec{q}^{\,})$ are the creation
and annihilation operators for a nucleon of momentum $\vec{q}$ while
$d^{\dagger}$ and $d$ are the corresponding anti-nucleon operators
and $u(\vec{q}^{\,})$ and $v(\vec{q}^{\,})$ are the positive and
negative energy  spinors corresponding to momentum $\vec{q}$,
normalized as in Itzykson and Zuber \cite{IZ84}, i.e. such that:
\begin {eqnarray} \overline{u}_\alpha(\vec{q}^{\,})
u_{\alpha'}(\vec{q}^{\,}) =\delta_{\alpha \alpha'}. \end {eqnarray}
Note that $E_N(\vec{q}^{\,})$ can be chosen to give non-relativistic,
semi-relativistic or fully relativistic kinematics. E.g., for full
relativistic kinematics choose:  \begin {equation}
E_N(\vec{q}^{\,})=\sqrt{m^2 + \vec{q}\,^2}. \end {equation} Since the
nucleons are fermions their operators obey the standard
anti-commutation relations:
\begin {eqnarray}
\{b_\alpha(\vec{q}^{\,}),b^{\dagger}_{\alpha'}(\vec{q}\,')\}=(2
\pi)^3 \frac{E_N(\vec{q}^{\,})}{m} \delta^{(3)}(q-q')\delta_{\alpha
\alpha'},\\
\{d_\alpha(\vec{q}^{\,}),d^{\dagger}_{\alpha'}(\vec{q}\,')\}=(2
\pi)^3 \frac{E_N(\vec{q}^{\,})}{m} \delta^{(3)}(q-q')\delta_{\alpha
\alpha'},
\end {eqnarray}
with all other anti-commutators vanishing.

Similarly, we define meson field operators $\phi_i(\vec{x})$, with
$i$ an isospin label, which we expand via:
\begin {equation}
\phi_i(\vec{x})=\int \frac{d^3k}{(2 \pi)^3 2 \omega_\pi(\vec{k})}
\left[ {a_i}(\vec{k}) e^{+i\vec{k} \cdot \vec{x}} + {a^\dagger_i}
(\vec{k}) e^{-i \vec{k} \cdot \vec{x}}) \right],  \label{eq:phi}
\end{equation} where $\omega_\pi(\vec{k})$ is the energy of a pion of
momentum $\vec{k}$, which in the relativistic kinematics is: \begin
{equation} \omega_\pi(\vec{k})=\sqrt{m_\pi^2 + \vec{k}^2}; \end
{equation} and ${a_i}(\vec{k})$ (${a^{\dagger}_i}(\vec{k})$) destroys
(creates) a pion with isospin label $i$ and momentum $\vec{k}$.
Again, this is in accordance with the definitions used by Itzykson
and Zuber \cite{IZ84}. The physical pion operators are then defined
by:
\begin {eqnarray}
\phi_{\pm}=\mp \frac{1}{\sqrt{2}} (\phi_1 \pm
i \phi_2),\\ \phi_0=\phi_3.
\end {eqnarray}
Note that while the
operators $\phi_i$ are self-adjoint the operators  $\phi_{\pm}$ are
not self-adjoint. Since pions are bosons the operators ${a_i}$ and
${a^{\dagger}_i}$ obey the commutation relations: \begin {eqnarray}
[a_i(\vec{k}),a^{\dagger}_{i'}(\vec{k}')]=2 \omega_\pi (\vec{k}) (2
\pi)^3 \delta^{(3)}(k-k') \delta_{i i'},\\ \end {eqnarray} with all
other commutators being zero.

These commutation relations are written for the Schr\"odinger
operators. Naturally, they remain the same if we change to some other
representation. The two other representations used in this paper are
the Heisenberg representation, which we shall denote by a tilde above
the operator in question, e.g. $\tilde{\psi},\tilde{\phi}$; and the
interaction representation, which we denote by a superscript $I$,
e.g. $\psi^I$, $\phi^I$. Schr\"odinger representation operators will
remain unadorned. These three different representations are connected
by the standard transformations involving the Hamiltonian of the
system.

We define this Hamiltonian  via the Lagrangian density, ${\cal L}$.
We choose:
\begin {equation} {\cal L}={\cal L_D} + {\cal L}_\phi
+{\cal L}_{int},
\end {equation}
where:
\begin {eqnarray} {\cal
L}_D(x)=\overline{\psi}(x) &(&i \gamma^\mu \partial_\mu - m)
\psi(x),\\ {\cal L}_\phi(x)=\frac{1}{2} (\partial_\mu \vec
{\phi}&(&x) \cdot \partial^\mu \vec{\phi}(x)  -  m_\pi^2
\vec{\phi}(x) \cdot \vec{\phi}(x)),
\end {eqnarray}
\begin {equation}
{\cal L}_{int}(x)=-ig \int d^4x_N \, d^4x_N' \, d^4x_\pi
\overline{\psi} (x_N') \gamma_5 \vec{\tau} \psi(x_N) \cdot
\vec{\phi}(x_\pi) \Gamma(x-x_N',x-x_N,x-x_\pi).
\end {equation} Here
we have chosen pseudo-scalar coupling for the $\pi-N$ interaction
Lagrangian density ${\cal L}_I(x)$, and
$\Gamma(x-x_N',x-x_N,x-x_\pi)$ is the  form factor for this
interaction. We introduce this form factor in order to model the
finite size of all the particles involved. It will also allow us to
remove the divergences from the theory. (See Figure \ref{fig-vertex}
for a depiction of this vertex.)

This Lagrangian is that used in the Walecka model (or Quantum
Hadrodynamics-1) \cite{Wa74,Se92} except for two differences: \begin
{enumerate} \item There are no counter-terms to do the
renormalization with. Instead, we include a form factor in the
pseudo-scalar interaction Lagrangian, thus allowing us to introduce a
cut-off and so remove the infinities from the theory.

\item In the QHD Lagrangian there are fields corresponding to one
family of vector mesons and one neutral scalar meson. The above
Lagrangian has only one family of pseudo-scalar mesons, the pions.
Note, however, that this Lagrangian could easily be extended to
include vector mesons.  \end {enumerate} The use to which this
Lagrangian is put will also be different. The QHD Lagrangian is used
for doing nuclear structure calculations and consequently is always
solved as a bound-state problem, often in the mean-field
approximation. Conversely, we will be using our Lagrangian to do
scattering theory---in particular, the scattering theory of the
$NN-\pi NN$ system.

{}From this Lagrangian, we may define the Hamiltonian density, ${\cal
H}$, by a Legendre transformation:\footnote{The canonical momenta,
$\pi$, for the field $\phi$, may not be given by the usual relation,
$\pi=\frac{\delta {\cal L}}{\delta \partial_0 \phi}$, due to the
presence of the dissipative force in the interaction Lagrangian. If
this is the case it means that the relation between the Lagrangian
and the Hamiltonian is more complicated than that given in Equation
(\ref{eq:LT}). We ignore this difficulty for the present, since we
are only really interested in the overall final form of the
Hamiltonian, which we may assume is given by Eqs.(\ref{eq:H}),
(\ref{eq:HK}) and (\ref{eq:HI}).}   \begin {equation} {\cal
H}(x)=\frac {\delta {\cal L}}{\delta \partial_0 \psi} \partial_0 \psi
+ \partial_0\overline{\psi} \frac {\delta {\cal L}}{\delta \partial_0
\overline{\psi}}  + \frac {\delta {\cal L}}{\delta \partial_0
\vec{\phi}} \cdot \partial_0 \vec {\phi} - {\cal L}   \label{eq:LT}
\end {equation}  and hence obtain the Hamiltonian:   \begin
{equation} H=\int d^3x \, {\cal H}(x)=H_K + H_{int}, \label{eq:H}
\end {equation} where: \begin {eqnarray} H_K&=&\int d^3x \,
\overline{\psi} (x) T_N(x) \psi(x) + \int d^3x \, \vec{\phi}(x)
T_\pi(x) \cdot \vec{\phi}(x), \label{eq:HK} \end {eqnarray} \begin
{equation} H_{int}=+ig \int d^3x \, (\int d^4x_N d^4x_N' d^4x_\pi
\overline{\psi} (x_N') \gamma_5 \vec{\tau} \psi(x_N) \cdot
\vec{\phi}(x_\pi) \Gamma(x-x_N',x-x_N,x-x_\pi)).  \label{eq:HI} \end
{equation} Note that we may define an interaction Hamiltonian density
\begin {equation} {\cal H}_{int}(x)=ig \int d^4x_N \, d^4x_N' \,
d^4x_\pi \overline{\psi} (x_N') \gamma_5 \vec{\tau} \psi(x_N) \cdot
\vec{\phi}(x_\pi) \Gamma(x-x_N',x-x_N,x-x_\pi).  \end {equation}

The connection between the position space operators of different
representations is now: \begin {eqnarray} \tilde{\psi}(x)&=&e^{i H
x^0} \psi(\vec{x}) e^{-i H x^0}\\ \psi^I(x)&=&e^{i H_K x^0}
\psi(\vec{x}) e^{-i H_K x^0}. \end {eqnarray} From these we may
deduce the results for $\overline{\psi}$. Furthermore, it is clear
that the same equations will hold if we replace $\psi$ by any
component of $\vec{\phi}$.

\section {The single-nucleon and single-pion Green's functions}

\label {sec-SingleptclGF}

In this section we first give a brief explanation of how to obtain a
diagrammatic expansion for the single-particle Green's functions, and
how to find the Feynman Rules in coordinate and momentum space for
the calculation of these Green's functions as a perturbation series
of Feynman diagrams. This is done primarily to fix normalizations and
phases. We then look at the form taken by the free Green's functions
in the low-energy limit, when anti-nucleonic contributions  to the
Green's function can be ignored.

We define the one-particle Green's functions for nucleons and pions
by: \begin {eqnarray} G_N(x',x) &=& \langle 0|T(\tilde{\psi}(x')
\tilde{\overline{\psi}}(x))|0 \rangle, \label{eq:gn0}\\
{G_\pi}_{ji}(x',x) &=& \langle 0|T(\tilde{\phi_j}(x')
\tilde{\phi_i}(x))|0 \rangle , \label{eq:gpi0}  \end {eqnarray}
respectively, where $T$ denotes the usual time-ordering operator.

Consider either of the one-particle Green's functions. By following
the standard procedure as outlined in e.g. Itzykson and Zuber
\cite{IZ84}, perturbation series for $G_N(x',x)$ and
${G_\pi}_{ji}(x',x)$ may be derived. We may then use Wick's theorem to
rewrite each term in these series as a sum of products of the free
single-particle Green's functions:  \begin {eqnarray} G_N^{(0)}
(x',x) &=& \langle 0|T (\psi^I(x') {\overline{\psi}}^I(x))|0
\rangle,\\ {G_\pi}^{(0)}_{ji}(x',x) &=& \langle 0|T (\phi_j^I(x')
\phi_i^I(x))|0 \rangle . \end {eqnarray} Consequently we are led to
the standard interpretation of the perturbation series as a series of
Feynman diagrams. Each Feynman diagram corresponds to an analytic
expression given by the Feynman Rules for the theory. The Feynman
Rules in coordinate space for the theory of nucleons and pions
obtained from the Hamiltonian of the previous Section are the standard
ones given in Bjorken and Drell, pp.224-5~\cite {BD64}, with two
changes: (i)~A form factor $\Gamma$ for pion absorption and emission
must be added, and (ii)~Rule 4 must be replaced by the following rule:

``4. Assign coordinates $x$ and $x'$ to the external points. If
dealing with ${G_\pi}_{ji}(x',x)$ also assign isospin indices $j$ and
$i$ to the external points. Then, if a nucleon line joins $x$ (or
$x'$) to an internal point $y$, associate with it a factor
$G_N^{(0)}(y,x)$ ($G_N^{(0)}(x',y)$). If a pion line joins $x$
($x'$), with isospin label $i$($j$), to a vertex $y$, with isospin
index $k$, then include a factor ${G_\pi}^{(0)}_{ki}(y,x)$
(${G_\pi}^{(0)}_{jk}(x',y)$) for that line."

Note that the change comes about because we are dealing with Green's
functions whereas Bjorken and Drell dealt with amplitudes.

Now we may define the Green's functions in momentum space, via:
\begin {equation} G_H(p',p)=\int d^4x \, d^4x' e^{ip'x'} G_H(x',x)
e^{-ipx}, \end {equation}  where $H$ is either a nucleon or a pion.
By translation invariance: \begin {equation} G_H(x',x)=G_H(x'-x) \end
{equation} and so it can be shown that: \begin {equation}
G_H(p',p)=(2 \pi)^4 \delta^{(4)} (p'-p) G_H(p), \end {equation}
where: \begin {equation} G_H(p)=\int d^4x_R \, e^{i p x_R} G_H(x_R).
\label {eq:pspacespGF} \end {equation} where $G_H(x_R)=G_H(x'-x)$ is
the coordinate space Green's function,  the rules for which are given
in Bjorken and Drell \cite{BD64}.

Consequently we may find the momentum space Green's function by first
using Wick's theorem to get an expansion for $G_H (x'-x)$ in terms of
the free nucleon and free pion Green's functions, and then inserting
the Fourier representations for the free Green's functions
$G_N^{(0)}(y'-y)$ and  ${G_\pi}^{(0)}_{ji}(y'-y)$ and the form factor
$\Gamma (y',y,y_\pi)$. For the free Green's functions we have:
\begin {eqnarray} G_N^{(0)}(y'-y)&=&\int \frac {d^4\tilde{q}}{(2
\pi)^4} e^{-i \tilde{q} (y'-y)} G_N^{(0)}(\tilde{q}),\\
{G_\pi}^{(0)}_{ji}(y'-y)&=&\int \frac {d^4k}{(2 \pi)^4} e^{-i k
(y'-y)} {G_\pi}^{(0)}_{ji}(k). \end {eqnarray} For the form factor we
define: \begin {equation} \Gamma(q',q,k)=\int d^4y \, d^4y' \,
d^4y_\pi e^{-i q' y'} e^{i k y_\pi} e^{i q y} \Gamma(y',y,y_\pi),
\label {eq:vertexFT} \end {equation} and the inverse of this Fourier
Transformation may be used to find the Fourier Representation of
$\Gamma(y',y,y_\pi)$.

The formulae thus obtained are substituted into the expression for a
particular Feynman diagram, the coordinate space integrations are
performed and finally the Fourier transform of the result is taken
using Eq.(\ref {eq:pspacespGF}). (For more detail on this procedure
see Itzykson and Zuber pp.267-8 \cite{IZ84}.)  When this is done the
following Feynman Rules for the construction of $G_H(p)$ are obtained.

\subsection*{Feynman Rules for the single-particle Green's function
in  momentum space}

\begin {enumerate}

\item Draw all topologically distinct, connected diagrams with one
incoming leg and one outgoing leg.

\hspace {-0.685 cm} In each diagram:

\item Assign to the external legs four-momentum $p$, and to the
internal nucleon and pion legs four-momenta $k_1,k_2,\ldots,k_I$,
where $I$ is the total number of internal lines. To each $\pi NN$
vertex assign an isospin label $l_1,l_2,\ldots,l_n$. If the external
legs are pion legs assign to the incoming (outgoing) pion leg's
endpoint an isospin label $i$ ($j$).

\item To the nucleon line with four-momentum $q$, where $q$ equals
some $k_j$, $j=1,\ldots,I$, assign a factor $G_N^{(0)}(q)$.

\item To the pion line with four-momentum $k$, connecting vertices
with isospin labels $l_e,l_a$, assign a factor ${G_\pi}^{(0)}_{l_a
l_e}(k)$.

\item To each pion emission vertex assign a factor: $$(-ig) (i
\gamma_5 \tau_{l}) (2 \pi)^4 \Gamma(q',q,-k) \delta^{(4)} (q'+k-q);$$
to each pion absorption vertex assign a factor: $$(-ig) (i \gamma_5
\tau_{l}) (2 \pi)^4 \Gamma(q',q,k) \delta^{(4)} (q'-q-k);$$ to each
$N \bar{N}$ annihilation vertex assign a factor: $$(-ig) (i \gamma_5
\tau_{l}) (2 \pi)^4 \Gamma(-q',q,-k) \delta^{(4)} (k-q-q');$$  and to
each $N \bar {N}$ creation vertex assign a factor: $$(-ig) (i
\gamma_5 \tau_{l}) (2 \pi)^4 \Gamma(q',-q,k) \delta^{(4)} (q'+q-k).$$
In each case $q$ ($q'$) is the nucleon four-momentum before (after)
the vertex (where applicable), $k$ is the pion four-momentum and $l$
is the isospin index assigned to the vertex.

\item Include a factor of (-1) for each closed fermion loop.

\item Integrate over all internal momenta: $$\frac{d^4k_1}{(2 \pi)^4}
\frac {d^4k_2}{(2 \pi)^4} \cdots \frac {d^4k_I}{(2 \pi)^4}$$  and sum
over all repeated isospin indices. \end {enumerate}

It is these rules we shall always work with in the calculations. But,
in order to calculate with these rules we need to know what the free
single-particle propagators $G_N^{(0)}(q)$ and $G_\pi^{(0)}(q)$ are.
It follows from Eqs.(\ref{eq:psi}), (\ref{eq:psib}), (\ref{eq:gn0})
and (\ref{eq:pspacespGF}) for the free single-nucleon Green's
function and Eqs.(\ref{eq:phi}), (\ref{eq:gpi0}),
(\ref{eq:pspacespGF}) for the free single-pion Green's function
that:   \begin {eqnarray}  G_N^{(0)}(q)&=&\frac{i}{\not\!{q} - m},
\label{eq:GNcov}\\ {G_{\pi}}_{ji}^{(0)}(k)&=&\frac{i
\delta_{ij}}{k^2-m_\pi^2}  \label{eq:Gpicov}, \end {eqnarray} i.e.
the free single-particle Green's functions are the full covariant
propagators. (See Appendix \ref{ap-GFcalc} for the details of this
calculation.)

\subsection*{The no-anti-nucleon approximation}

However, the use of these full covariant propagators results, in most
cases, in a very difficult calculation.  In order to simplify the
calculation somewhat, a restriction is placed upon Green's functions
in this work. We make the approximation that the contribution made by
anti-nucleons to the Green's functions is negligible, an
approximation which should be true in the low energy limit. Making
this approximation means that we force:    \begin {equation}
v(\vec{q}^{\,}) \rightarrow 0;\overline{v}(\vec{q}^{\,}) \rightarrow
0. \end {equation} We expect this to be true at the energies we are
interested in. It immediately implies that: \begin {equation}
G_N^{(0)} (x',x)= \langle 0|\psi^I(x') {\overline{\psi}}^I(x)|0
\rangle  \theta(x_0' - x_0), \label{eq:thetaG} \end {equation} as one
would expect in the absence of anti-nucleons. It is also found that,
as a direct consequence of the derivation of the single-nucleon
Green's function in momentum space, we have: \begin {equation}
G_N^{(0)}(q)=\frac{im}{E_N(\vec{q}^{\,})} \frac {\sum_\alpha u_\alpha
(\vec{q}^{\,}) \overline{u}_\alpha(\vec{q}^{\,})}{q_0^+ -
E_N(\vec{q}^{\,})}, \label {eq:pfreeNGF} \end {equation} in this
approximation. (See Appendix \ref{ap-GFcalc} for the justification of
these two statements.)

This no-anti-nucleon approximation not only affects the free
single-nucleon propagator, it imposes a definite time-ordering on any
process contributing to the single-nucleon Green's function. This
occurs because the theta-functions in the free single-nucleon
propagators define the order of any two pion emission and absorption
times $y_0$ and $y_0'$. I.e., the absence of anti-nucleons determines
which of the two times in any free-pion Green's function,
${G_\pi}^{(0)}_{ji} (y',y)$, comes first and which second. So, we
have to replace ${G_\pi}^{(0)}_{ji} (y',y)$ by ${G_\pi}^{(0)}_{ji}
(y',y) \theta(y_0' - y_0)$, where $y_0'$ is the later of the two
times. In momentum space this has the effect of replacing
Eq.(\ref{eq:Gpicov}) by:    \begin {equation}
{G_\pi}^{(0)}_{ji}(k)=\frac{1}{2 \omega_\pi(\vec{k})} \frac{i
\delta_{ij}}{k_0^+ - \omega_\pi(\vec{k})}. \label{eq:pfreepiGF} \end
{equation} (Again, a justification of this fact is to be found in
Appendix \ref{ap-GFcalc}.)

Furthermore, in the approximation in which anti-nucleons cannot be
created, it follows from the conservation of nucleon number that no
processes other than free pion propagation can contribute to the
single-pion Green's function. Therefore:   \begin {equation}
{G_\pi}_{ji}(k)={G_\pi}_{ji}^{(0)}(k). \end {equation}

By contrast, if we attempt to express $G_N$ in terms of $G_N^{(0)}$
we find that $G_N$ obeys a Schwinger-Dyson equation.

So, the no-anti-nucleon approximation reduces the momentum space
propagators from full covariant propagators to those given in
Eqs.(\ref{eq:pfreeNGF}) and (\ref{eq:pfreepiGF}) and prevents any
process other than free propagation from contributing to the
single-pion Green's function.

\section {The two-nucleon Green's function}

\label {sec-TwonucleonGF}

In the previous section we explained how to obtain a diagrammatic
expansion and Feynman Rules in both coordinate and momentum space for
the single-nucleon and single-pion Green's functions. Since our aim
is to develop a theory of the $NN-\pi NN$ system we should next
consider the $NN$ Green's function and its relation to $G_N$ and
${G_\pi}_{ji}$.

The two-nucleon Green's function is defined by: \begin {equation}
G_{NN}(x_1',x_2';x_1,x_2)= \langle 0|T(\tilde{\psi}_{N}(x_1')
\tilde{\psi}_{N}(x_2') \tilde {\overline{\psi}}_{N}(x_1) \tilde
{\overline{\psi}}_{N}(x_2))|0 \rangle . \label{eq:GNNdef} \end
{equation} The procedure for obtaining the Feynman rules for this
Green's function is exactly the same as that used for the
single-particle Green's functions in the previous section. Upon
implementing that procedure we find that the rules obtained in
coordinate space for the two-nucleon Green's function are the same as
those for the one-nucleon Green' function, with two modifications:
(i)~In the two-nucleon case diagrams must obviously have two incoming
and two outgoing external nucleon lines, and (ii)~Disconnected
diagrams are now permitted, provided they contain no sub-diagrams
which are not ultimately joined to the external lines. (For examples
of the type of diagrams which are and are not allowed see Figure
\ref{fig-vacvac}.)

Now we attempt to obtain the Feynman Rules in momentum space for the
two-nucleon Green's function, but with a restriction that the initial
times on both particles are equal, as are the final times. It turns
out that this restriction of equal times generates a form for the
Green's function bearing a remarkable similarity to the convolution
integral used in the work of Kvinikhidze and Blankleider (KB)
\cite{BK92A,BK92B}. The argument we give is valid regardless of
whether or not we are using the no-anti-nucleon approximation.

Consider the two-nucleon Green's function
$G_{NN}(x_1,x_2;x_1',x_2')$. Suppose that we restrict the times the
particles begin and end their propagation, so  that the two nucleons
begin their propagation at equal times, i.e.:  \begin {equation}
x_1^0=x_2^0 \equiv t, \end {equation} and end it at equal times, i.e.:
\begin {equation} {x_1^0}'={x_2^0}' \equiv t'. \end {equation} Then
we denote the Green's function by: \begin {equation}
\overline{G}_{NN} (t',\vec{x}_1^{\, \prime},\vec{x}_2^{\,
\prime};t,\vec{x}_1,\vec{x}_2). \end {equation}

Now consider the Fourier Transform of this Green's function. Because
there is only one initial and one final time, when the Fourier
Transform with respect to energy is taken, instead of transforming
with respect to the individual particle variables $p_1^0$, $p_2^0$,
${p_1^0}'$ and ${p_2^0}'$, which are conjugate to $x_1^0$, $x_2^0$,
${x_1^0}'$ and ${x_2^0}'$, we must transform with respect to
variables $E$ and $E'$ which are conjugate to $t$ and $t'$.
Therefore,   \begin {eqnarray}  & &\overline{G}_{NN}
(E',\vec{p}_1^{\, \prime},\vec{p}_2^{\,
\prime};E,\vec{p}_1,\vec{p}_2)\nonumber\\ & &=\int dt \, dt'
 d^3x_1  d^3x_2  d^3x_1' d^3x_2' e^{i(E't' - \vec{p}_1^{\, \prime}
\cdot \vec{x}_1^{\, \prime} - \vec{p}_2^{\, \prime} \cdot
\vec{x}_2^{\, \prime})} G_{NN} (t',\vec{x}_1^{\,
\prime},\vec{x}_2^{\, \prime};t,\vec{x}_1,\vec{x}_2) e^{-i(Et -
\vec{p}_1 \cdot \vec{x}_1 - \vec{p}_2 \cdot \vec{x}_2)}\\ & &= \int
d^4x_1 \, d^4x_2 \, d^4x_1' \, d^4x_2' e^{i(E'{x_1'}^0 -
\vec{p}_1^{\, \prime} \cdot \vec{x}_1^{\, \prime} - \vec{p}_2^{\,
\prime} \cdot \vec{x}_2^{\, \prime})} \delta ({x_1^0}' -{x_2^0}')
G_{NN}(x_1',x_2';x_1,x_2) \delta (x_1^0 - x_2^0) \nonumber\\ & &  \:
\times e^{-i(Ex_1^0 - \vec{p}_1 \cdot \vec{x}_1 - \vec{p}_2 \cdot
\vec{x}_2)}, \end {eqnarray}  where we have recalled that $x_1^0
\equiv t$ and ${x_1^0}' \equiv t'$ and then used the two-time Green's
function $G_{NN}(x_1',x_2';x_1,x_2)$ in place of the equal-time
Green's function. Inserting:   \begin {equation} \delta(a-b)=\int
\frac{dz}{2 \pi} e^{iz(a-b)}, \end {equation} and then performing the
Fourier transforms over $x_1,x_2,x_1'$ and $x_2'$ gives: \begin
{equation} \overline{G}_{NN}(E',\vec{p}_1^{\, \prime},\vec{p}_2^{\,
\prime};E,\vec{p}_1,\vec{p}_2)=\int \frac {dz \, dz'}{(2 \pi)^2}
G_{NN}(E'-z',\vec{p}_1^{\, \prime},z',\vec{p}_2^{\,
\prime};E-z,\vec{p}_1,z,\vec{p}_2), \label {eq:CI1}  \end {equation}
where $G_{NN}({p_1^0}',\vec{p}_1^{\, \prime},{p_2^0}',\vec{p}_2^{\,
\prime};p_1^0,\vec{p}_1,p_2^0,\vec{p}_2)$ is the Fourier transform of
$G_{NN}(x_1',x_2';x_1,x_2)$. At this stage it becomes clear that $E$
and $E'$ are the initial and final total energy of the system,
respectively: \begin {eqnarray} E &=& p_1^0 + p_2^0,\\ E' &=&
{p_1^0}' + {p_2^0}'. \end {eqnarray} It can also be seen that the
integral over $z$ and $z'$ corresponds to an integral over all
possible initial and final "relative" energies of the two-nucleon
system.\footnote {Note that the definition of "relative" energy used
here differs from the standard $k^0=p_1^0-p_2^0$. Instead, when the
term relative energy is used in this paper it refers to the energy
$z=p_2^0=E-p_1^0$.}

Now the translational invariance of  $G_{NN}(x_1',x_2';x_1,x_2)$
implies momentum conservation in $G_{NN}(p_1',p_2';p_1,p_2)$, i.e.
\begin {equation} G_{NN}(p_1',p_2';p_1,p_2)=(2 \pi)^4 \delta^{(4)}
(p_1' + p_2' - p_1 - p_2) G_{NN}(p_1',p_2';p_1,p_2). \end {equation}
This suggests that: \begin {eqnarray} G_{NN}(E'-z',\vec{p}_1^{\,
\prime},z',\vec{p}_2^{\, \prime};E&-&z,\vec{p}_1,z,\vec{p}_2)\\ = (2
\pi)^4 \delta(E'-E) \delta^{(3)} (p_1' &+& p_2' - p_1 - p_2)
G_{NN}(E-z',\vec{p}_1^{\, \prime},z',\vec{p}_2^{\,
\prime};E-z,\vec{p}_1,z,\vec{p}_2). \end {eqnarray} Further,
translational invariance of $\overline{G}_{NN}(t',\vec{x}_1^{\,
\prime},\vec{x}_2^{\, \prime};t,\vec{x}_1,\vec{x}_2)$ implies: \begin
{equation} \overline{G}_{NN}(E',\vec{p}_1^{\, \prime},\vec{p}_2^{\,
\prime};E,\vec{p}_1,\vec{p}_2)=(2 \pi)^4 \delta (E'-E) \delta^{(3)}
(p_1' + p_2' - p_1 - p_2) \overline{G}_{NN}(E,\vec{p}_1^{\,
\prime},\vec{p}_2^{\, \prime},\vec{p}_1,\vec{p}_2).  \label
{eq:Epcons} \end {equation} These last two results suggest that
Eq.(\ref{eq:CI1}) may be rewritten as: \begin {equation}
\overline{G}_{NN}(E,\vec{p}_1^{\, \prime},\vec{p}_2^{\,
\prime},\vec{p}_1,\vec{p}_2)=\int \frac{dz \, dz'}{(2 \pi)^2}
G_{NN}(E-z',\vec{p}_1^{\, \prime},z',\vec{p}_2^{\,
\prime};E-z,\vec{p}_1,z,\vec{p}_2) \label {eq:CI2}. \end {equation}
Therefore the problem of calculating $G_{NN}(E,\vec{p}_1^{\,
\prime},\vec{p}_2^{\, \prime},\vec{p}_1,\vec{p}_2)$ is reduced to
that of calculating $G_{NN}(p_1',p_2';p_1,p_2)$. But
$G_{NN}(p_1',p_2';p_1,p_2)$ is to be calculated using the Feynman
Rules in momentum space for the two-nucleon Green's function. To
obtain these Feynman rules from those in coordinate space we merely
proceed as we did above for the one-nucleon Green's function. This
leads to Feynman Rules for the calculation of
$G_{NN}(E-z',\vec{p}_1^{\, \prime},z',\vec{p}_2^{\,
\prime};E-z,\vec{p}_1,z,\vec{p}_2)$ which are exactly the same as
those for the single-particle Green's function, except that Rules 1
and 2 become:

\begin {enumerate} \item Draw all topologically distinct diagrams
with two incoming and two outgoing legs, excluding diagrams in which
any sub-diagram is not ultimately connected to an external line.

\hspace {-0.685 cm} For each diagram:

\item Assign to the incoming (outgoing) external nucleon legs
energies $E-z$ and $z$ ($E-z'$ and $z'$) and momenta $\vec{p}_1$ and
$\vec{p}_2$ ($\vec{p}_1^{\, \prime}$ and $\vec{p}_2^{\, \prime}$).
Assign to the internal nucleon and pion legs momenta
$k_1,\ldots,k_I$, where $I$ is the number of internal lines. Note
that once the momenta and energy of the particles in the initial
state are chosen there are two ways of choosing the momenta and
energy of the particles in the final state. Both possible assignments
must be included as separate diagrams. The two resulting diagrams
differ by a factor of (-1), since the nucleons are fermions. \end
{enumerate}

This is the only change which needs to be made to the full covariant
theory in which anti-nucleons are included. However, if we are using
the no-anti-nucleon approximation another change is needed in order
to maintain consistency in our work, as follows. Recall that, in the
calculation of the single-nucleon Green's function we had to replace
${G_\pi}^{(0)}_{ji} (y'-y)$ by ${G_\pi}^{(0)}_{ji} (y'-y) \theta(y_0'
- y_0)$, because the no-anti-nucleon approximation imposed a definite
time-order on pion propagation. However, in the two-nucleon Green's
function there is nothing which specifies the relative time-order of
processes taking place on different nucleons. Therefore, pions
emitted on one nucleon and absorbed on the other {\em do not} have
the time-order of their emission and absorption specified, whereas,
as described above, pions emitted and absorbed on the same nucleon
{\em do} have that time-order specified. So, in order to treat
transmitted pions and pions that are reabsorbed by the same nucleon
which emits them equivalently, we split the transmitted pion's
Green's function ${G_\pi}^{(0)}_{ji} (y_2,y_1)$ into two
possibilities, according to whether $y_1^0 < y_2^0$ or $y_2^0 <
y_1^0$, and insert appropriate theta functions. Therefore, for
transmitted pions one diagram in which we have
${G_\pi}^{(0)}_{ji}(y',y)$is replaced  by two diagrams, in one of
which we have ${G_\pi}^{(0)}_{ji}(y',y) \theta(y_0' - y_0)$, and in
the other of which we have ${G_\pi}^{(0)}_{ji}(y',y) \theta(y_0 -
y_0')$. This rule ensures that the relative time-order of a
particular pion's emission and absorption is always determined. It is
important to note that, other than this restriction, the relative
time-orders of processes on different nucleons remains
undetermined.

If we use the no-anti-nucleon approximation and make this change in
order to maintain consistency then we still have the same rules for
the two-nucleon Green's function as in the full covariant theory, but
with the simplified propagators given in Eqs.(\ref{eq:pfreeNGF}) and
(\ref{eq:pfreepiGF}).

\section {The multi-particle Green's function}

\label {sec-mm'ptclGF}

The argument for the $m \rightarrow m'$ particle Green's function is
constructed exactly as are the above arguments for the one and
two-nucleon Green's functions. The only change necessary to the
Feynman rules for the two-nucleon Green's function in coordinate
space  is that we must now draw all possible topologically distinct
diagrams with $m$ incoming and $m'$ outgoing legs.

Now we examine the equal-time Green's function, in order to derive a
result analogous to Equation (\ref{eq:CI1}). The argument used is
similar to that by which Eq.(\ref{eq:CI1}) was derived.

The Fourier transform of the equal-time $m \rightarrow m'$ particle
Green's function: \begin {equation} \overline{G}(t',\vec{x}_1^{\,
\prime},\vec{x}_2^{\, \prime},\ldots,\vec{x}_{m'}^{\,
\prime};t,\vec{x}_1,\vec{x}_2,\ldots,\vec{x}_m), \end {equation}
which we denote by $\overline{G}(E',\vec{p}_1^{\,
\prime},\vec{p}_2^{\, \prime},\ldots,\vec{p}_{m'}^{\,
\prime};E,\vec{p}_1,\vec{p}_2,\ldots,\vec{p}_m)$, is given by the
formula: \begin {eqnarray*} & & \overline{G}(E',\vec{p}_1^{\,
\prime},\vec{p}_2^{\, \prime},\ldots,\vec{p}_{m'}^{\,
\prime};E,\vec{p}_1,\vec{p}_2,\ldots,\vec{p}_m)=\int \frac {dz^{(2)}
dz^{(3)} \cdots dz^{(m)} d{z^{(2)}}' d{z^{(3)}}' \cdots
d{z^{(m')}}'}{(2 \pi)^{m+m'-2}}\\  & & \times \mbox{}
G(E'-\sum_{i=2}^{m'} {z^{(i)}}',\vec{p}_1^{\,
\prime},{z^{(2)}}',\vec{p}_2^{\,
\prime},\ldots,{z^{(m')}}',\vec{p}_{m'}^{\, \prime};E-\sum_{i=2}^m
z^{(i)},\vec{p}_1,z^{(2)},\vec{p}_2,\ldots,z^{(m)},\vec{p}_m), \end
{eqnarray*} \noindent where
$G(p_1',p_2',\ldots,p_{m'}';p_1,p_2,\ldots,p_m)$ is to be calculated
using the Feynman rules in momentum space for the $m\rightarrow m'$
particle Green's function. These are the Feynman rules for the
two-nucleon Green's function, with changes to Rules 1 and 2 to
accommodate the different numbers of particles in the initial and
final states. In Rule 1 we replace "two incoming and two outgoing" by
"$m$ incoming and $m'$ outgoing" and Rule 2 becomes:

"2. Assign to the incoming external particle legs four-momenta
$p_1,p_2,\ldots,p_m$; assign to the outgoing legs four-momenta
$p_1',p_2',\ldots,p_{m'}'$. Assign to the internal nucleon and pion
legs momenta $k_1,\ldots,k_I$, where $I$ is the number of internal
lines. Note that once the four-momenta of the particles in the
initial state are chosen there are $m'!$ ways of choosing the
four-momenta of the particles in the final state. All possible
assignments must be included, as separate diagrams, with appropriate
relative signs. If, for one such diagram, we make an assignment of
the fermion four-momenta $(p_1',\ldots,p_{n_F}')$, where $n_F$ is the
number of fermions in the final state, then the relative sign of the
other $m'! - 1$ diagrams generated from it by permutation of the
fermion four-momenta is (+1) for an even permutation of
$(p_1',\ldots,p_{n_F}')$ and (-1) for an odd permutation."

Therefore, the equal-time $m\rightarrow m'$ single-energy Green's
function: $$\overline{G}(E,\vec{p}_1^{\, \prime},\vec{p}_2^{\,
\prime},\ldots,\vec{p}_{m'}^{\,
\prime},\vec{p}_1,\vec{p}_2,\ldots,\vec{p}_m)$$ may be found by
calculating  $$G(E'-\sum_{i=2}^{m'} {z^{(i)}}',\vec{p}_1^{\,
\prime},{z^{(2)}}',\vec{p}_2^{\,
\prime},\ldots,{z^{(m')}}',\vec{p}_{m'}^{\, \prime};E-\sum_{i=2}^m
z^{(i)},\vec{p}_1,z^{(2)},\vec{p}_2,\ldots,z^{(m)},\vec{p}_m)$$
according to these rules and integrating over
$z^{(2)},z^{(3)},\ldots,z^{(m)},{z^{(2)}}',\ldots,\!{z^{(m')}}'$.
Again, this result is true regardless of whether or not the
anti-nucleon  approximation is used or not.

\section {Derivation of old-fashioned time-ordered perturbation
theory and the result of Kvinikhidze and Blankleider from this
perturbation scheme}

\label {sec-BKconnection}

In this section we will show how the above formalism allows the
summation of all relative time-orders in an old-fashioned
time-ordered perturbation theory diagram, and so derive the result of
Kvinikhidze and Blankleider (KB) \cite{BK92A}. In the light of this
calculation we will then comment on certain approximations made by KB
in their work on the $NN-\pi NN$ problem \cite{BK92B}. In order to
achieve these goals it is necessary to first examine how to derive
old-fashioned perturbation theory rules for the two-nucleon Green's
function from the above Feynman rules. It will be clear from the
structure of this derivation that we could generalize this argument
to the $m \rightarrow m'$ Green's function.

\subsection {Derivation of Feynman rules for time-ordered
perturbation theory}

Consider the equal-time Green's function in the no-anti-nucleon
approximation:  \begin {equation} \overline{G}_{NN} (t',\vec{x}_1^{\,
\prime},\vec{x}_2^{\, \prime};t,\vec{x}_1,\vec{x}_2). \end {equation}
Define equal-time two-nucleon annihilation and creation operators in
the Heisenberg representation via: \begin {eqnarray}
\tilde{\psi_{NN}}(t,\vec{x}_1,\vec{x}_2)&=&\tilde{\psi}(t,\vec{x}_1)
\tilde{\psi} (t,\vec{x}_2),\\
\tilde{\overline{\psi}_{NN}}(t,\vec{x}_1,\vec{x}_2)&=&\tilde{\overline{\psi}}
(t,\vec{x}_2) \tilde{\overline{\psi}} (t,\vec{x}_1).  \end {eqnarray}
The interaction and Schr\"odinger representation equal-time
two-nucleon operators may then be found in the usual way.

Currently, the perturbation expansion for $G_{NN}(t',\vec{x}_1^{\,
\prime},\vec{x}_2^{\, \prime};t,\vec{x}_1,\vec{x}_2)$ is written in
terms of individual particle operators, i.e. $\psi^I$s,
${\overline{\psi}}^I$s and $\phi^I_i$s. The first step in our
derivation of time-ordered perturbation theory is to rewrite this
perturbation expansion in terms of the two-nucleon operator,
$\psi_{NN}^I$, and the pionic operators, $\phi_i^I$. The only change
which needs to be made in order to rewrite the perturbation expansion
in this way is to prove that the $\pi NN$ vertex:  \begin {equation}
{\cal H}_{int}(x)=ig \int d^4x_N \, d^4x_N' \, d^4x_\pi
\overline{\psi} (x_N') \gamma_5 \vec{\tau} \psi(x_N) \cdot
\vec{\phi}(x_\pi) \Gamma(x-x_N',x-x_N,x-x_\pi), \end {equation} which
we currently use, may be replaced by a vertex: \begin {eqnarray}
&{{\cal H}_{int}}_{(2)}(x)&=ig \int d^3x_{ }' (\int d^4x_N \, d^4x_N'
\, d^4x_\pi  \overline{\psi}_{NN}({x_N^0}',\vec{x}_N^{\,
\prime},\vec{x}^{\, \prime}) \gamma_5 \vec{\tau} \nonumber\\ & &
\times \psi_{NN}(x_N^0,\vec{x}_N,\vec{x}^{\, \prime}) \cdot
\vec{\phi}(x_\pi) \Gamma(x-x_N',x-x_N,x-x_\pi)) \end {eqnarray} in
which not a single nucleon, but a nucleon pair, is destroyed and
recreated.  (See Figure \ref{fig-newvertex} for a diagrammatic
representation of this new vertex.) The proof that, in the absence of
anti-nucleons, rewriting the vertex in this way does not change the
value of the Green's function is outlined in Appendix
\ref{ap-changeonetotwo}. This result then allows us to rewrite the
expansion for $\overline{G}_{NN}(t',\vec{x}_1^{\,
\prime},\vec{x}_2^{\, \prime};t,\vec{x}_1,\vec{x}_2)$ obtained from
Wick's theorem in terms of the free Green's functions:   \begin
{eqnarray} \overline{G}_{NN}^{(0)}(t',\vec{x}_1^{\,
\prime},\vec{x}_2^{\, \prime};t,\vec{x}_1,\vec{x}_2)&=&
 \langle 0|T(\psi^I_{NN}(t',\vec{x}_1^{\, \prime},\vec{x}_2^{\,
\prime}) \overline{\psi}^I_{NN}(t,\vec{x}_1,\vec{x}_2))|0 \rangle, \\
{G_{\pi}}^{(0)}_{ji}(y',y)&=& \langle 0|T(\phi_j^I(y') \phi_i^I(y))|0
\rangle .   \end {eqnarray}  When we do this we obtain a set of
Feynman Rules in coordinate space for this theory.

Now we examine the consequences of making the no-anti-nucleon
approximation. The first consequence of this approximation is that,
similarly to the one-nucleon case,  \begin {equation}
\overline{G}_{NN}^{(0)}(t',\vec{x}_1^{\, \prime},\vec{x}_2^{\,
\prime};t,\vec{x}_1,\vec{x}_2)=0 \mbox { if $t > t'$}. \end {equation}
Since in the no-anti-nucleon approximation this Green's function must
appear between any two  vertices it follows that the time-order of
{\em all} vertices is now determined. This is to be contrasted with
the perturbation scheme used in the previous section where the
relative time-order of most of the events occurring on different
nucleons was {\em not} determined. In that perturbation scheme
absorption and emission of exchanged pions were the only events
occurring on different nucleons whose relative time-order was
specified. In this perturbation scheme the relative time-order of all
events is determined.

Other than this change, however, the two perturbation schemes result
in the same Green's function, since the rewriting of the equal-time
two-nucleon Green's function in terms of free two-nucleon Green's
functions does not change its value. Suppose then that we call the
result of a particular diagram evaluated according to the rules of the
previous Section:  \begin {equation} {\overline{G}_{NN}}_{(1)}
(t',\vec{x}_1^{\, \prime},\vec{x}_2^{\,
\prime};t,\vec{x}_1,\vec{x}_2), \end {equation} where the subscript
$(1)$ indicates the use of one-nucleon Green's functions in the
calculation. Suppose also that we call the result constructed by the
rules obtained below for the theory with a two-nucleon vertex: \begin
{equation} {\overline{G}_{NN}}_{(2)} (t',\vec{x}_1^{\,
\prime},\vec{x}_2^{\, \prime};t,\vec{x}_1,\vec{x}_2), \end {equation}
where the subscript $(2)$ indicates the use of two-nucleon Green's
functions in this calculation. Then it follows that:  \begin
{equation} \sum_{\mbox {All relative TOs}}{\overline{G}_{NN}}_{(2)}
(t',\vec{x}_1^{\, \prime},\vec{x}_2^{\,
\prime};t,\vec{x}_1,\vec{x}_2)  ={\overline{G}_{NN}}_{(1)}
(t',\vec{x}_1^{\, \prime},\vec{x}_2^{\,
\prime};t,\vec{x}_1,\vec{x}_2), \label {eq:TOPTsum} \end {equation}
where "All relative TOs" indicates that the sum is over all possible
relative time-orders of events on different nucleons, with the
exception of the emission and absorption of exchanged pions, whose
relative time-order is fixed as soon as the diagram to be considered
is chosen. This result may also be deduced from Equation
(\ref{eq:GNNsum}) which was derived in Appendix
\ref{ap-changeonetotwo}. Note that both Green's functions must be
evaluated in the no-anti-nucleon approximation if this equation is to
hold. We shall return to Eq.(\ref{eq:TOPTsum}) shortly.

First, however, we must complete our development of time-ordered
perturbation theory, for we have not yet explained how the rules
usually used in the calculation of ${\overline{G}_{NN}}_{(2)}
(t',\vec{x}_1^{\, \prime},\vec{x}_2^{\,
\prime};t,\vec{x}_1,\vec{x}_2)$ come about.

Take the Fourier Transform of the Green's function:
$${\overline{G}_{NN}}_{(2)} (t',\vec{x}_1^{\, \prime},\vec{x}_2^{\,
\prime};t,\vec{x}_1,\vec{x}_2),$$ in order to obtain:
$${\overline{G}_{NN}}_{(2)}(E',\vec{p}_1^{\, \prime},\vec{p}_2^{\,
\prime};E,\vec{p}_1,\vec{p}_2).$$ Just as was done for
${\overline{G}_{NN}}_{(1)}$ above, in Eq.(\ref{eq:Epcons}),
energy-momentum conservation in this Green's function may be
expressed via: \begin {equation}
{\overline{G}_{NN}}_{(2)}(E',\vec{p}_1^{\, \prime},\vec{p}_2^{\,
\prime};E,\vec{p}_1,\vec{p}_2)=(2 \pi)^4 \delta (E'-E) \delta^{(3)}
(p_1' + p_2' - p_1 - p_2) {\overline{G}_{NN}}_{(2)}(E,\vec{p}_1^{\,
\prime},\vec{p}_2^{\, \prime},\vec{p}_1,\vec{p}_2).  \end {equation}
The rules for calculating ${\overline{G}_{NN}}_{(2)}(E,\vec{p}_1^{\,
\prime},\vec{p}_2^{\, \prime},\vec{p}_1,\vec{p}_2)$ are found from
the rules in co-ordinate space by the standard procedure explained in
Section \ref{sec-SingleptclGF}. The rules thus obtained are as
follows:

\subsection*{Feynman Rules for the two-nucleon Green's function in
terms of the free two-nucleon Green's function, in the
no-anti-nucleon approximation.} \begin {enumerate}

\item Draw all those topologically distinct, connected diagrams which
have two incoming and two outgoing legs. Remember that different
time-orders now contribute to different diagrams.

\hspace {-0.685 cm} In each diagram:

\item Label the pion legs with momenta $k_1,\ldots,k_{I'}$, label the
external nucleon pairs with combined energy-momenta
$(E,\vec{p}_1,\vec{p}_2)$ and $(E',\vec{p}_1^{\,
\prime},\vec{p}_2^{\, \prime})$. There are two possible ways of
assigning the momenta $\vec{p}_1^{\, \prime}$ and $\vec{p}_2^{\,
\prime}$ in the final state, and both possibilities should be
included as separate diagrams, with a relative minus sign. Label the
internal nucleon pairs with combined energy-momenta:
$$(E^{(1)},\vec{q}_1^{\, (1)},\vec{q}_2^{\, (1)}),
(E^{(2)},\vec{q}_1^{\, (2)},\vec{q}_2^{\, (2)}),\ldots,
(E^{(n-1)},\vec{q}_1^{\, (n-1)},\vec{q}_2^{\, (n-1)}).$$  Assign
isospin labels $l_1,l_2,\ldots,l_n$ to the $n$ vertices. Note that
because of our rewriting of ${\cal H}_{int}$ above, a new nucleon
pair is regarded as being created whenever a vertex occurs on {\em
either} nucleon.

\item To the nucleon pair with combined energy-momentum
$(E,\vec{q}_1,\vec{q}_2)$ assign a factor
${\overline{G}}^{(0)}_{NN}(E,\vec{q}_1,\vec{q}_2)$.

\item To the pion line with four-momentum $k$, joining vertices with
isospin labels $l_e$ and $l_a$, assign a factor ${G_\pi}_{l_a
l_e}^{(0)}(k)$.

\item For each pion absorption vertex occurring on the nucleon with
momentum $\vec{q}_i$ insert a factor:  \begin {eqnarray} (-ig) (2
\pi)^7 \delta &(& E'-k_0-E) \delta^{(3)}(q_i'-k-q_i)
\delta^{(3)}(q_{\bar {i}}' - q_{\bar {i}}^{}) (i \gamma_5
\tau_{l})\nonumber\\ \times \Gamma
&(&E'-E_N(\vec{q}_{\bar{i}}^{\,\prime}),\vec{q}_i^{\,\prime},
E-E_N(\vec{q}_{\bar{i}}),\vec{q_i},k_0,\vec{k}).  \nonumber  \end
{eqnarray}  For each emission vertex occurring under the same
assumptions insert a similar factor, but with $(k_0,\vec{k})$
replaced by $(-k_0,-\vec{k})$.

Here  \begin {equation}  \bar{i}=\left \{ \begin{array}{ll}
 2 & \mbox { if $i=1$}\\ 1 & \mbox { if $i=2$.} \end{array}  \right.
\end {equation} Also, $(E,\vec{q}_i,\vec{q}_{\bar{i}})$ is the
combined energy-momentum of the nucleon-pair before the pion emission
or absorption; $(E',\vec{q}_i^{\, \prime},\vec{q}_{\bar{i}}^{\,
\prime})$ is the combined energy-momentum of the nucleon-pair
afterwards, $(k_0,\vec{k})$ is the four-momentum of the pion and $l$
is the isospin label assigned to the vertex.

\item Integrate over all internal energies and momenta: $$\frac
{dE^{(1)}d^3q_1^{(1)}d^3q_2^{(1)}}{(2 \pi)^7} \frac
{dE^{(2)}d^3q_1^{(2)}d^3q_2^{(2)}}{(2 \pi)^7} \cdots \frac
{dE^{(n-1)}d^3q_1^{(n-1)}d^3q_2^{(n-1)}}{(2 \pi)^7} \frac{d^4k_1}{(2
\pi)^4} \frac {d^4k_2}{(2 \pi)^4} \cdots \frac {d^4k_{I'}}{(2
\pi)^4}$$  and sum over all repeated isospin indices. \end {enumerate}

In these rules ${\overline{G}}_{NN}^{(0)}(E,\vec{q}_1,\vec{q}_2)$ is
defined by: \begin {eqnarray} (2 \pi)^7 \delta(E'-E)
\delta^{(3)}(q_1'-q_1) \delta^{(3)}(q_2'-q_2)
{\overline{G}}_{NN}^{(0)}(E,\vec{q}_1,\vec{q}_2)&=&\int dt \, dt' \,
d^3x_1 \, d^3x_2 \, d^3x_1' \, d^3x_2' \nonumber\\  \times e^{i(E't'
- \vec{p}_1^{\, \prime} \cdot \vec{x}_1^{\, \prime} - \vec{p}_2^{\,
\prime} \cdot \vec{x}_2^{\, \prime})}
{\overline{G}}_{NN}^{(0)}(t',\vec{x}_1^{\, \prime},\vec{x}_2^{\,
\prime};t,\vec{x}_1,\vec{x}_2) \theta(t'&-&t) e^{-i(Et - \vec{p}_1
\cdot \vec{x}_1 - \vec{p}_2 \cdot \vec{x}_2)}. \end {eqnarray} When
this Fourier transform is evaluated we find: \begin {equation}
{\overline{G}}_{NN}^{(0)}(E,\vec{q}_1,\vec{q}_2)=\frac{im^2}{E_N(\vec{q}_1)
E_N(\vec{q}_2)} \frac {\sum_{\alpha_1,\alpha_2} u_{\alpha_1}
(\vec{q}_1) \overline {u}_{\alpha_1}(\vec{q}_1) u_{\alpha_2}
(\vec{q}_2) \overline {u}_{\alpha_2}(\vec{q}_2)}{E^+ - E_N(\vec{q}_1)
- E_N(\vec{q}_2)}. \end {equation} At the same time
${G_\pi}^{(0)}_{ji}(k)$ is defined exactly as in previous sections
and so, in this no-anti-nucleon approximation, ${G_\pi}_{ji}^{(0)}(k)$
is given by Eq.(\ref{eq:pfreepiGF}).

Note that all the arguments in this section assume that the vertex
$\Gamma$ has no analytic structure in its energy variables. This
assumption is justified since the presence of analytic structure
would indicate that intermediate states in $\Gamma$ could be exposed.
Because $\Gamma$ is the bare vertex for the theory it cannot have
such intermediate states, and so it must have no analytic structure
in its energy variables.

This represents one way of formulating the Feynman Rules for the
two-nucleon Green's function in this "time-ordered" or "old-fashioned"
perturbation theory.

There is another more usual way of writing these rules. Instead of
representing  a $j$-pion intermediate state using one Green's
function for the two nucleons  and another $j$ for the pions, we may
write a single Green's function for all of the $j+2$ particles in
this state. This change can be accomplished either at the co-ordinate
space level (in a similar but more complex way to the above change
from an expression in terms of one-nucleon operators to one in terms
of two-nucleon operators), or it can be done at the momentum-space
level, as follows.

Consider any diagram containing a $j$-pion intermediate state.
Suppose that this diagram is evaluated using the above Feynman
rules.  We know from these rules that, if the isospin indices which
are irrelevant to our argument are suppressed, the Green's function
for the diagram may be written:  \begin {eqnarray}
{\overline{G}}^{j}& &(E',\vec{p}_1^{\, \prime},\vec{p}_2^{\,
\prime};E,\vec{p}_1,\vec{p}_2)=\nonumber\\ \int & & \frac {d\tilde{E}
d^3{\tilde{p_1}} d^3{\tilde{p_2}}}{(2 \pi)^7} \frac{d^4k^{(1)}
d^4k^{(2)} \cdots d^4k^{(j)}}{(2 \pi)^{4j}} F (E',\vec{p}_1^{\,
\prime},\vec{p}_2^{\,
\prime};\tilde{E},\tilde{\vec{p}_1},\tilde{\vec{p}_2},k^{(1)},
k^{(2)} ,\ldots,k^{(j)})
{\overline{G}}^{(0)}_{NN}(\tilde{E},\tilde{\vec{p}_1},\tilde{\vec{p}_2})
\nonumber\\ & & \times G_\pi^{(0)}(k^{(1)}) G_\pi^{(0)} (k^{(2)})
\cdots G_\pi^{(0)}(k^{(j)})
F^{\dagger}(\tilde{E},\tilde{\vec{p}_1},\tilde{\vec{p}_2},k^{(1)},
k^{(2)},\ldots,k^{(j)};E,\vec{p}_1,\vec{p}_2).   \end {eqnarray}
\noindent (See Figure \ref{fig-jpionint}.) The Feynman Rules also
ensure that energy is conserved in $F$ and $F^\dagger$. In
particular, we may extract a factor $(2 \pi)
\delta(E-\tilde{E}-\sum_{m=1}^j k_0^{(m)})$ from $F^\dagger$ and a
factor $(2 \pi) \delta(\tilde{E}+\sum_{m=1}^j k_0^{(m)}-E')$ from
$F$. Furthermore, $F$ and $F^\dagger$ have no poles in any of their
$j$ $k_0$-variables, because the propagator for the $m$th pion, which
is the only part of the expression to contain analytic structure in
the variable $k_0^{(m)}$, appears explicitly in our equation for
${\overline{G}}^j$ and so cannot form part of $F$ and $F^\dagger$.

Using this energy-conservation result and substituting for
${\overline{G}}_{NN}^{(0)}$ and $G_\pi^{(0)}$ gives: \begin {eqnarray}
{\overline{G}}^j(E',\vec{p}_1^{\, \prime},\vec{p}_2^{\,
\prime}&;&E,\vec{p}_1,\vec{p}_2)\nonumber\\* = (2 \pi) \delta&(&E'-E)
\int \frac {d^3\tilde{p_1} d^3\tilde{p_2} d^3k^{(1)} d^3k^{(2)}
\cdots d^3k^{(j)}}{(2 \pi)^{3j + 6}} \frac {dk_0^{(1)} dk_0^{(2)}
\cdots dk_0^{(j)}}{(2\pi)^j} \nonumber\\* & & F(E',\vec{p}_1^{\,
\prime},\vec{p}_2^{\, \prime};E - \sum_{m=1}^j
k_0^{(m)},\tilde{\vec{p}_1},\tilde{\vec{p}_2},k^{(1)}, k^{(2)},
\ldots, k^{(j)})\nonumber\\* \frac{im^2}{E_N(\tilde{\vec{p}_1})
E_N(\tilde{\vec{p}_2})} & & \frac {\sum_{\alpha_1,\alpha_2}
u_{\alpha_1} (\tilde{\vec{p}_1}) \overline
{u}_{\alpha_1}(\tilde{\vec{p}_1}) u_{\alpha_2} (\tilde{\vec{p}_2})
\overline {u}_{\alpha_2}(\tilde{\vec{p}_2})}{E^+ - \sum_{m=1}^j
k_0^{(m)} - E_N (\tilde{\vec{p}_1}) - E_N(\tilde{\vec{p}_2})}
\prod_{m=1}^j \frac{1}{2 \omega_\pi(\vec{k}^{(m)})} \,
\frac{i}{{k_0^{(m)}}^+ - \omega_\pi(\vec{k}^{(m)})}\nonumber\\ & &
F^\dagger (E-\sum_{m=1}^j
k_0^{(m)},\tilde{\vec{p}_1},\tilde{\vec{p}_2},k^{(1)}, k^{(2)},
\ldots, k^{(j)};E,\vec{p}_1,\vec{p}_2). \end {eqnarray} Performing
the $k_0$ integrations then gives: \begin {eqnarray}
{\overline{G}}^j(E',\vec{p}_1^{\, \prime},\vec{p}_2^{\,
\prime};E,\vec{p}_1,\vec{p}_2)=(2 \pi) & & \delta(E'-E) \int \frac
{d^3\tilde{p_1} d^3\tilde{p_2} d^3k^{(1)} d^3k^{(2)} \cdots
d^3k^{(j)}}{(2 \pi)^{3j + 6}}\nonumber\\  F(E',\vec{p}_1^{\,
\prime},\vec{p}_2^{\, \prime};E - \sum_{m=1}^j
\omega_\pi(\vec{k}^{(m)}),\tilde{\vec{p}_1}, \tilde{\vec{p}_2},& &
\omega_\pi(\vec{k}^{(1)}),\vec{k}^{(1)},\omega_\pi(\vec{k}^{(2)}),
\vec{k}^{(2)}, \ldots,\omega_\pi(\vec{k}^{(j)}) ,\vec{k}^{(j)})
\nonumber\\ \frac{im^2}{E_N(\tilde{\vec{p}_1})
E_N(\tilde{\vec{p}_2})} \prod_{m=1}^j \frac{1}{2
\omega_\pi(\vec{k}^{(m)})}  & & \frac {\sum_{\alpha_1,\alpha_2}
u_{\alpha_1} (\tilde{\vec{p}_1}) \overline
{u}_{\alpha_1}(\tilde{\vec{p}_1}) u_{\alpha_2} (\tilde{\vec{p}_2})
\overline {u}_{\alpha_2}(\tilde{\vec{p}_2})}{E^+ - \sum_{m=1}^j
\omega_\pi (\vec{k}^{(m)}) - E_N (\tilde{\vec{p}_1}) -
E_N(\tilde{\vec{p}_2})}\nonumber\\  F^{\dagger} (E-\sum_{m=1}^j
\omega_\pi (\vec{k}^{(m)})
,\tilde{\vec{p}_1},\tilde{\vec{p}_2};\omega_\pi(\vec{k}^{(1)}),& &
\vec{k}^{(1)},\omega_\pi(\vec{k}^{(2)}),k^{(2)},\ldots,\omega_\pi(\vec{k}^{(j)}),\vec{k}^{(j)};E,\vec{p}_1,\vec{p}_2).
\end {eqnarray}

Since this argument is valid for any $j$-pion intermediate state the
above set of Feynman rules may be replaced by the following set of
Feynman rules, which are those  usually used for calculating the
Green's function in old-fashioned time-ordered perturbation theory.

\subsection*{Feynman rules for old-fashioned time-ordered
perturbation theory}

\begin {enumerate}

\item As for Rule 1 above.

\hspace {-0.685 cm} In each diagram:

\item Label the incoming (outgoing) nucleon pairs with momenta
$(\vec{p}_1,\vec{p}_2)$ ($(\vec{p}_1^{\, \prime},\vec{p}_2^{\,
\prime})$) and internal nucleon pairs with momenta
$(\vec{q}_1^{\,(1)},\vec{q}_2^{\,(1)}),
(\vec{q}_1^{\,(2)},\vec{q}_2^{\,(2)}),\ldots,
(\vec{q}_1^{\,(n-1)},\vec{q}_2^{\,(n-1)})$, including separate
diagrams with a relative minus sign for the two possible momentum
assignments in the final state. Label internal pion lines with
momenta $\vec{k}^{(1)},\vec{k}^{(2)},\ldots,\vec{k}^{(I')}$. Assign
isospin labels $l_1,l_2,\ldots,l_n$ to the vertices. Recall that a
new nucleon pair is regarded as being created whenever a vertex
occurs on {\em either} nucleon.

\item If $j$ pions are present at the same time as the nucleon pair,
and they have momenta $\vec{k}^{(1)},\ldots,\vec{k}^{(j)}$ while the
nucleon pair has momenta $\vec{q}_1$ and $\vec{q}_2$, then associate,
with all the particles present at that time, a Green's function:
\begin {equation} \frac{im^2}{E_N(\vec{q}_1)  E_N(\vec{q}_2)}
\prod_{m=1}^j \frac{1}{2 \omega_\pi(\vec{k}^{(m)})}
\frac{\sum_{\alpha_1,\alpha_2} u_{\alpha_1} (\vec{q}_1) \overline
{u}_{\alpha_1}(\vec{q}_1) u_{\alpha_2} (\vec{q}_2) \overline
{u}_{\alpha_2}(\vec{q}_2)}{E^+-E_N(\vec{q}_1) - E_N(\vec{q}_2) -
\sum_{m=1}^j \omega_\pi (\vec{k}^{(m)})} \prod_{l=1}^{j} \delta_{l_e
l_a},  \end {equation} where $l_e$ and $l_a$ are the isospin indices
of the vertex at which the $l$th pion is emitted and absorbed.

Here "$j$ pions present at the same time" means that a vertical line
drawn through the two nucleon lines intersects $j$ pion lines.

\item For each pion absorption vertex occurring on the nucleon with
momentum $\vec{q}_i$ at the same time as $j$ spectator pions with
momenta $\vec{k}_1,\ldots,\vec{k}_j$ are present, insert a factor:
\begin {eqnarray} (-ig) (2 \pi)^3  \delta^{(3)}&(&q_i'-k-q_i) (2
\pi)^3 \delta^{(3)}(q_{\bar {i}}' - q_{\bar {i}})  (2 \pi)^{3j}
\prod_{m=1}^j \delta^{(3)}(k_m'-k_m) (i \gamma_5 \tau_{l})\nonumber\\
\times \Gamma &(&E - \sum_{m=1}^j \omega_\pi(\vec{k}_{j}') -
E_N(\vec{q}_{\bar{i}}^{\, \prime}), \vec{q}_i^{\, \prime},E -
E_N(\vec{q}_{\bar{i}}),\omega_\pi(\vec{k}),\vec{k}). \nonumber
\end{eqnarray}  Here $l$ is the isospin label assigned to the vertex.

For an emission vertex occurring under the same conditions we insert
a similar factor, but with $\vec{k}$ replaced by $-\vec{k}$ and
$\omega_\pi(\vec{k})$ replaced by $-\omega_\pi (\vec{k})$.

\item Integrate over all internal momenta: $$\frac{d^3k_1}{(2 \pi)^3}
\frac {d^3k_2}{(2 \pi)^3} \cdots \frac {d^3k_{I'}}{(2 \pi)^3} \frac
{d^3q_1^{(1)} d^3q_2^{(1)}}{(2\pi)^6} \frac {d^3q_1^{(2)}
d^3q_2^{(2)}}{(2\pi)^6} \cdots \frac {d^3q_1^{(n-1)}
d^3q_2^{(n-1)}}{(2\pi)^6}$$    and sum over all repeated isospin
indices. \end {enumerate}

\subsection {The result of Kvinikhidze and Blankleider}

Having obtained the usual time-ordered perturbation theory rules for
calculating ${\overline{G}_{NN}}_{(2)}(E',\vec{p}_1^{\,
\prime},\vec{p}_2^{\, \prime};E,\vec{p}_1,\vec{p}_2)$ we may use
Eq.(\ref{eq:TOPTsum}) to derive some of the results proven by KB
using different means \cite{BK92A,BK92B,BK93}.

Consider any time-ordered perturbation theory diagram, and suppose
that the Green's function for the diagram is
${\overline{G}_{NN}}_{(2)}$. Define a Green's function
${\overline{G}^{\Sigma}_{NN}}_{(2)}$ which is the sum of all Green's
functions representing  diagrams which differ from the diagram under
consideration only in the relative time-order of processes occurring
on different nucleons, but excluding those diagrams in which the
order of emission and absorption of transmitted pions is different.
I.e. define ${\overline{G}^\Sigma_{NN}}_{(2)}$ to be:   \begin
{equation} {\overline{G}^{\Sigma}_{NN}}_{(2)}=\sum_{\mbox {All
relative time-orders}} {\overline{G}_{NN}}_{(2)}. \end {equation}

Once this definition is made we may take the Fourier transform of
Eq.(\ref{eq:TOPTsum}) and use Eq.(\ref{eq:CI1}) to obtain: \begin
{eqnarray} {\overline{G}^{\Sigma}_{NN}}_{(2)}(E',\vec{p}_1^{\,
\prime},\vec{p}_2^{\, \prime};E,\vec{p}_1,\vec{p}_2&)&=\nonumber\\
\int \frac{dz dz'}{(2\pi)^2} {G_{NN}}_{(1)} &(&E'-z',\vec{p}_1^{\,
\prime},z',\vec{p}_2^{\, \prime};E-z,\vec{p}_1,z,\vec{p}_2),   \label
{eq:BKdelta}   \end {eqnarray} where ${G_{NN}}_{(1)}$ must be
evaluated in the no-anti-nucleon approximation. Or, if we remove the
energy delta-function on both sides: \begin {equation}
{\overline{G}^{\Sigma}_{NN}}_{(2)}(E,\vec{p}_1^{\,
\prime},\vec{p}_2^{\, \prime},\vec{p}_1,\vec{p}_2)=\int \frac{dz
dz'}{(2\pi)^2} {G_{NN}}_{(1)} (E-z',\vec{p}_1^{\,
\prime},z',\vec{p}_2^{\, \prime};E-z,\vec{p}_1,z,\vec{p}_2).
\label{eq:BKnodelta}  \end {equation}  These are two of the crucial
results of this paper. They state that the Green's function for a sum
of time-ordered perturbation theory diagrams may be expressed as the
integral over all initial and final relative energies of another
Green's function, which is simply the covariant Green's function in
the no-anti-nucleon approximation. This is the general result, now
let us look at two specific examples. Firstly, we will consider an
arbitrary disconnected diagram and show how Eq.(\ref{eq:BKdelta})
reduces to the original KB convolution integral result for this case.
Secondly, we will examine one-pion exchange as an example of what
happens when we apply Eq.(\ref{eq:BKnodelta})  to a connected diagram.

\subsubsection{Disconnected diagrams}

Consider an arbitrary disconnected diagram. By definition, no pion is
transmitted from one nucleon to the other. Suppose that, in this
diagram, processes $P1$ occur on one nucleon and processes $P2$ occur
on the other. Applying the Feynman Rules in Section
\ref{sec-TwonucleonGF} to this situation we see that:  \begin
{eqnarray} {G_{NN}^{(d)}}_1&(&E'-z',\vec{p}_1^{\,
\prime},z',\vec{p}_2^{\, \prime};E-z,\vec{p}_1,z,\vec{p}_2)=  (2
\pi)^8 \delta (z'-z) \delta (E'-E) \nonumber\\  & & \times
\delta^{(3)}(p_1' - p_1) \delta^{(3)}(p_2' - p_2)
G_N^{P1}(E-z,\vec{p}_1) G_N^{P2}(z,\vec{p}_2), \end {eqnarray} where
$G_N^{P1}(E-z,\vec{p}_1)$ and $G_N^{P2}(z,\vec{p}_2)$ are the
single-nucleon Green's functions for the processes $P1$ and $P2$
respectively, and are to be calculated from the rules in Section
\ref{sec-SingleptclGF}, applied in the no-anti-nucleon approximation.

Compare this with the form of ${\overline{G}_{NN}}_{(2)}$ for the
same diagram: \begin {equation} {\overline{G}^{\Sigma \,
(d)}_{NN}}_{(2)} (E',\vec{p}_1^{\, \prime},\vec{p}_2^{\,
\prime};E,\vec{p}_1,\vec{p}_2)=(2 \pi)^7 \delta (E' - E)
\delta^{(3)}(p_1' - p_1) \delta^{(3)}(p_2' - p_2)
{{\overline{G}^{\Sigma}_{NN}}^{(d)}_2} (E,\vec{p}_1,\vec{p}_2). \end
{equation} Applying Eq.(\ref{eq:BKnodelta}) to this specific case
consequently yields: \begin {equation} {\overline{G}^{\Sigma \,
(d)}_{NN}}_{(2)} (E,\vec{p}_1,\vec{p}_2) =\int \frac{dz}{2 \pi}
G_N^{P1}(E-z,\vec{p}_1) G_N^{P2}(z,\vec{p}_2), \label{eq:BKdisc} \end
{equation} which is KB's result up to a factor of $i$ \cite{BK92A}.
This factor is missing because we are using propagators which differ
from KB's propagators by a factor of $i$. If we replace all our
propagators by KB propagators then, for an $n$th order diagram, a
factor $i^{n+1}$ appears on the left-hand side of
Eq.(\ref{eq:BKdisc}) and a factor $i^{n+2}$ appears on the right-hand
side. Therefore, we find that, for any diagram:  \begin {equation}
{\overline{G}^{\Sigma \, (d)}_{NN}}_{(2)} (E,\vec{p}_1,\vec{p}_2)
=-\frac{1}{2 \pi i} \int dz \, G_N^{P1}(E-z,\vec{p}_1)
G_N^{P2}(z,\vec{p}_2), \label{eq:BKres} \end {equation} where the
propagators now agree with KB's propagators.

\subsubsection {One-pion exchange}

Now we consider one-pion exchange. Suppose a pion is transmitted from
one nucleon, which we label $N2$, to the other, which we label $N1$.
Suppose that before the pion is emitted from $N2$ (absorbed on $N1$)
certain processes $P1$ ($P2$) take place on $N1$ ($N2$), and after
the pion is emitted on $N2$ (absorbed on $N1$) certain processes
$P1'$ ($P2'$) take place on $N1$ ($N2$). Then the Green's function
constructed according to the rules in Section \ref{sec-TwonucleonGF}
is:  \begin {eqnarray} {G_{NN}}_{(1)}(E-z',\vec{p}_1^{\,
\prime},z',\vec{p}_2^{\, \prime};E-z,\vec{p}_1,z,\vec{p}_2)= (2 \pi)^3
\delta^{(3)}(p_1' + p_2' - p_1 - p_2) \nonumber\\       \times
G_N^{P1'}(E-z',\vec{p}_1^{\, \prime}) (-ig) \Gamma(E-z',\vec{p}_1^{\,
\prime},E-z,\vec{p}_1,z-z',\vec{p}_1^{\, \prime}-\vec{p}_1)
G_N^{P1}(E-z,\vec{p}_1)\nonumber\\  \times
G_\pi(z-z',\vec{p}_2-\vec{p}_2^{\, \prime})  G_N^{P2'}
(z',\vec{p}_2^{\, \prime}) (-ig)   \Gamma(z',\vec{p}_2^{\,
\prime},z,\vec{p}_2,z'-z,\vec{p}_2^{\, \prime}-\vec{p}_2) G_N^{P2}
(z,\vec{p}_2),     \end {eqnarray}    where all single-particle
propagators $G_N^P (p_0,\vec{p}^{\,})$ are calculated using the rules
giving the single-particle Green's function for a process $P$, given
in Section \ref{sec-SingleptclGF}, applied in the no-anti-nucleon
approximation, and we have suppressed all the spin and isospin
structure of the result.

Substituting this into Eq.(\ref{eq:BKnodelta}) leads to:  \begin
{eqnarray} {\overline{G}_{NN}^\Sigma}_{(2)} (E,\vec{p}_1^{\,
\prime},\vec{p}_2^{\, \prime},\vec{p}_1,\vec{p}_2)=(2 \pi)^3
\delta^{(3)}(p_1'+p_2'-p_1-p_2) \int \frac {dz \, dz'}{(2
\pi)^2}\nonumber\\ \times G_N^{P1'}(E-z',\vec{p}_1^{\, \prime}) (-ig)
\Gamma(E-z',\vec{p}_1^{\, \prime},E-z,\vec{p}_1,z-z',\vec{p}_1^{\,
\prime}-\vec{p}_1)
 G_N^{P1} (E-z,\vec{p}_1)\nonumber\\ \times G_\pi
(z-z',\vec{p}_2-\vec{p}_2^{\, \prime}) G_N^{P2'} (z',\vec{p}_2^{\,
\prime}) (-ig)  \Gamma(z',\vec{p}_2^{\,
\prime},z,\vec{p}_2,z'-z,\vec{p}_2^{\, \prime}-\vec{p}_2)   G_N^{P2}
(z,\vec{p}_2). \end {eqnarray}    If we again remove the relative
factor of $i$ between our propagators and those of KB we obtain, for
an $n$th order diagram,  a factor $i^{n+2}$ on the right-hand side
and $i^n$ on the left-hand side. Therefore, for any order diagram the
result is:   \begin {eqnarray}
{\overline{G}^\Sigma_{NN}}_{(2)}(E,\vec{p}_1^{\,
\prime},\vec{p}_2^{\, \prime},\vec{p}_1,\vec{p}_2)=
\delta^{(3)}(p_1'+p_2'-p_1-p_2) (-\frac {1}{2 \pi i})^2 \int  dz \,
dz' \nonumber\\  \times G_N^{P1'}(E-z',\vec{p}_1^{\, \prime})  (-ig)
\Gamma(E-z',\vec{p}_1^{\, \prime},E-z,\vec{p}_1,z-z',\vec{p}_1^{\,
\prime}-\vec{p}_1)  G_N^{P1} (E-z,\vec{p}_1) \nonumber\\  \times
{G_\pi} (z-z',\vec{p}_2-\vec{p}_2^{\, \prime}) G_N^{P2'}
(z',\vec{p}_2^{\, \prime}) (-ig)  \Gamma(z',\vec{p}_2^{\,
\prime},z,\vec{p}_2,z'-z,\vec{p}_2^{\, \prime}-\vec{p}_2)
G_N^{P2}(z,\vec{p}_2), \label {eq:CIOPE}  \end {eqnarray} where, once
again, the propagators now agree with KB's propagators. If one now
sums this result over all possible processes $P1$, $P2$, $P1'$ and
$P2'$ one obtains the sum of all time-orders of one-pion exchange
from $N2$ to $N1$ with fully dressed propagators.  \begin {eqnarray}
{\overline{G}_{NN}^\Sigma}_{(2)} (E,\vec{p}_1^{\,
\prime},\vec{p}_2^{\,
\prime},\vec{p}_1,\vec{p}_2)=\delta^{(3)}(p_1'+p_2'-p_1-p_2) (-\frac
{1}{2 \pi i})^2 \int  dz dz' \nonumber\\  \times
G_N(E-z',\vec{p}_1^{\, \prime}) (-ig) \Gamma(E-z',\vec{p}_1^{\,
\prime},E-z,\vec{p}_1,z-z',\vec{p}_1^{\, \prime}-\vec{p}_1) G_N
(E-z,\vec{p}_1)\nonumber\\   \times {G_\pi}
(z-z',\vec{p}_2-\vec{p}_2^{\, \prime}) G_N (z',\vec{p}_2^{\, \prime})
(-ig) \Gamma(z',\vec{p}_2^{\, \prime},z,\vec{p}_2,z'-z,\vec{p}_2^{\,
\prime}-\vec{p}_2)  G_N (z,\vec{p}_2),  \label {eq:CIdOPE}  \end
{eqnarray} where all of the propagators $G_N$ are fully dressed. This
is another result also obtained by KB, as an extension of their
formula (\ref{eq:BKres}) \cite{BK93}. As we shall see in Section
\ref{sec-KLMK} this formula for the one-pion exchange potential was
also obtained by Klein and McCormick \cite{KM58}. (See Figure
\ref{fig-FullCIOPE} for a pictorial representation of this expression
and the energy and momentum assignments involved.)

Clearly, this approach can be extended to two pion exchange, three
pion exchange etc.. The advantage of such a method of calculating
Green's functions is that it allows us to sum all relative
time-orders in time-ordered perturbation theory by merely doing a
convolution integral of the result obtained from the rules given in
Section \ref{sec-TwonucleonGF}, applied in the no-anti-nucleon
approximation.

\subsection {The approximation of Kvinikhidze and Blankleider}

\label {sec-KBapprox}

Equation (\ref{eq:CIdOPE}) is the exact result for the sum of all
relative time-orders for one-pion exchange from $N2$ to $N1$, with
fully dressed propagators. However, in their work on the $NN-\pi NN$
problem KB use only an approximation to this result \cite{BK92B}.
They represent the one-pion exchange Green's function by a product of
two convolution integrals separated by an inverse $\pi NN$
propagator:   \begin {eqnarray} {\overline{G}_{NN}^\Sigma}_{(2)}
(E,\vec{p}_1^{\, \prime},\vec{p}_2^{\, \prime},\vec{p}_1,\vec{p}_2)
\approx \delta^{(3)}(p_1'+p_2'-p_1-p_2) [(-\frac {1}{2 \pi i})^2
\int dz' dz''  \nonumber\\* \times G_N (E-z',\vec{p}_1^{\, \prime})
(-ig) \Gamma(E-z',\vec{p}_1^{\,
\prime},E-z'',\vec{p}_1,z''-z',\vec{p}_2-\vec{p}_2^{\, \prime})
\nonumber\\* \times G_{N} (E-z'',\vec{p}_1)
{G_{\pi}}(z''-z',\vec{p}_1^{\, \prime}-\vec{p}_1) G_{N}
(z',\vec{p}_2^{\, \prime}) ] [{G_{\pi NN}}^{-1}(E)] [(-\frac {1}{2
\pi i})^2 \int dz''' dz \nonumber\\* \times G_{N} (z''',\vec{p}_2^{\,
\prime}) {G_{\pi}} (z-z''',\vec{p}_2-\vec{p}_2^{\, \prime}) (-ig)
\Gamma(z''',\vec{p}_2^{\, \prime},z,\vec{p}_2,z'''-z,\vec{p}_2^{\,
\prime}-\vec{p}_2) \nonumber\\* \times G_N (E-z,\vec{p}_1)   G_N
(z,\vec{p}_2)], \label{eq:aBKOPE}    \end {eqnarray} \noindent where,
once again, all propagators are fully dressed. KB have performed
calculations which suggest that, at least for this case, the error in
making such an approximation is small \cite{BK93}. Be that as it may,
we believe that such an approximation is physically flawed on at
least two grounds, and consequently we expect it to produce invalid
results, if not in the calculation of one-pion exchange, then in some
other calculation.

Firstly, to try and express one-pion exchange as in
Eq.(\ref{eq:aBKOPE}) violates conservation of energy in the following
sense. If the energy arguments of the Green's functions are treated
as off-shell energies of the particles involved, then the energy and
momentum assignments in Eq.(\ref{eq:aBKOPE}) correspond to the
energies and momenta indicated in Figure \ref{fig-BKapOPE}. These
momenta are to be compared to those in the exact expression for the
sum of time-ordered perturbation theory diagrams, depicted in Figure
\ref{fig-FullCIOPE}. It is clear that in the exact expression the
energy and momentum of each individual particle is conserved
throughout the diagram, except at the vertices, where a pion
interacts with a nucleon, thus modifying the energy and momentum of
the nucleon with which it interacts. However, the $\pi NN$ vertices
are constructed in such a way that energy and momentum are conserved
there too; therefore, one can say that the energy and momentum of the
individual particles is conserved in this exact expression for the
sum of time-ordered perturbation theory diagrams. In the approximate
expression, the momentum of each individual particle is still
conserved, in the same sense. However, an inspection of the energy
assignments reveals that, although the total energy is the same at
all times during the pion exchange process, the energy of the three
individual particles jumps suddenly and for no physical reason at the
point where the two disconnected pieces are joined together to form a
connected diagram. Since the relative energy of the particles is
integrated over this jump in energy is possibly quite large, and so
this violation of energy conservation is bound to have an effect
somewhere in any calculation performed using such an approximation.

A second associated flaw in such an approximation lies in the
diagrams it omits from the Green's function. The types of diagrams
omitted are represented in Figure \ref{fig-BKOmit}. The justification
for ignoring such diagrams is that, as mentioned above, KB have
performed a calculation which shows their effect to be small
\cite{BK93}. This appears reasonable, until one carefully examines
what physics is omitted and what is included in such an
approximation. Observe that diagrams such as Figures \ref{fig-BKInc}
and \ref{fig-BKInc2} are still included in KB's calculation. Breaking
down these two included diagrams and the omitted diagram into smaller
pieces implies that KB's approximation amounts to {\em always}
including the process on the left of Figure \ref{fig-BKJenn} but only
{\em sometimes} including the process on the right. The process on
the right of Figure \ref{fig-BKJenn} is excluded in some diagrams,
such as Figure \ref{fig-BKOmit}, but included in others, such as
Figure \ref{fig-BKInc2}. In other words, in  certain diagrams KB
ignore the Jennings mechanism \cite{Je88,JR88} when a spectator pion
is present. I.e., diagrams which contain the same physical processes
are treated differently, with some time-orders included and some
excluded.

These two problems reflect a fundamental flaw in the KB approach. KB
use a convolution integral in order to sum all relative time-orders
in a disconnected diagram, an approach, which, as we have seen, leads
to the calculation of a Green's function which is a no-anti-nucleon
approximation to the full covariant propagator. In other words, in
KB's model, disconnected pieces are calculated in an approach {\em
which is an approximation to covariant perturbation theory}. However,
when they come to construct convolution integral expressions for
connected diagrams they do so by joining disconnected pieces, written
as convolution integrals, together using an inverse $\pi NN$
propagator, as in Equation (\ref{eq:aBKOPE}). This method of joining
disconnected diagrams together is consistent with {\em time-ordered
perturbation theory}, not with covariant perturbation theory.
Consequently, their technique for forming connected diagrams from
disconnected pieces is inconsistent with their calculation of
disconnected diagrams. This leads, firstly, to violations of energy
conservation, since although the total energy $E$ is conserved the
relative energy is {\em not} constrained to be conserved, and
secondly, to the unequal treatment of certain diagrams which only
differ in the relative time-order of processes on different nucleons,
which is, ironically, the problem the convolution integral was
introduced to fix.

If one wishes to join disconnected pieces together in a fashion
consistent with the use of a convolution integral for disconnected
diagrams one needs to reexamine the roots of the convolution integral
technique, as we have done here. This leads to expressions such as
Eq.(\ref{eq:CIOPE}) which are the {\em exact} sums of all relative
time-orders in a set of time-ordered perturbation theory diagrams.
But, such a result is only an equal-time no-anti-nucleon
approximation to the full covariant perturbation theory expression.
Therefore, we intend to persist with the covariant calculation in
which all the physics is naturally included, rather than pursuing
approximations with questionable physical motivation, such as those
used by Kvinikhidze and Blankleider.
\section {Amplitudes}

\label {sec-LSZredn}

In this section we obtain the amplitudes which correspond to
the two-nucleon Green's functions we have used so far. This is
done via the standard procedure for obtaining amplitudes from
Green's functions: LSZ reduction \cite{LSZ55}, in which the
external legs of the Green's function are amputated in order
to obtain the corresponding  amplitude. Again, it will be
clear that the work of this section can be extended to any $m
\rightarrow m'$ particle Green's function.

For the two-nucleon system the S-matrix is defined to be:
\begin {eqnarray}
 \langle p_1',p_2',\mbox{out}|p_1,p_2,\mbox{in} \rangle &=& (2
\pi)^8 \delta^{(4)} (p_1' - p_1) \delta^{(4)} (p_2'-p_2)
\nonumber\\ + (2 \pi)^4 \delta^{(4)} (p_1' + p_2' &-& p_1 -
p_2)  \langle p_1',p_2'|T_{NN}|p_1,p_2 \rangle,  \end
{eqnarray} where: \begin {equation}
 \langle p_1',p_2'|T_{NN}|p_1,p_2 \rangle =G_N^{-1}(p_1')
G_N^{-1}(p_2') G_{NN}(p_1',p_2';p_1,p_2) G_N^{-1}(p_1)
G_N^{-1}(p_2), \end {equation} and we have ignored spin and
isospin indices. Note the use of dressed nucleon propagators
$G_N$ here.\footnote {Since we are using dressed nucleon
propagators we should include factors of $Z^{-\frac{1}{2}}$,
where $Z$ is the wave function renormalization, if we wish to
get the LSZ reduction formula exactly correct. However, the
inclusion of such factors makes no difference to the overall
result and so we omit them in order to make the argument as
clear as possible.} Firstly, we apply this formula to the
general two-nucleon Green's function obtained from the rules
in Section \ref{sec-TwonucleonGF}:  \begin {equation}
{G_{NN}}_{(1)}(E-z',\vec{p}_1^{\, \prime},z',\vec{p}_2^{\,
\prime};E-z,\vec{p}_1,z,\vec{p}_2). \end {equation} Note that
we have defined a total energy
$E={p_1^0}'+{p_2^0}'=p_1^0+p_2^0$ and relative energies
$z'={p_2^0}'$,$z=p_2^0$ in order to facilitate the derivation
of a relation between the amplitude obtained from this Green's
function and the one obtained from the time-ordered
perturbation theory Green's function. When the free
one-nucleon Green's function is substituted in, this equation
becomes, in the no-anti-nucleon approximation:  \begin
{eqnarray} {G_{NN}}_{(1)}(E-z',\vec{p}_1^{\,
\prime},z',\vec{p}_2^{\, \prime};E-z,\vec{p}_1,z,\vec{p}_2)=
\frac{i}{E^+-z'-E_N(\vec{p}_1^{\, \prime})}
\frac{i}{{z'}^+-E_N(\vec{p}_2^{\, \prime})}\nonumber\\  \times
{T_{NN}}_{(1)} (E-z',\vec{p}_1^{\, \prime},z',\vec{p}_2^{\,
\prime};E-z,\vec{p}_1,z,\vec{p}_2)
\frac{i}{E^+-z-E_N(\vec{p}_1)} \frac{i}{z^+-E_N(\vec{p}_2)},
\label {eq:LSZSPGF}   \end {eqnarray} where $E_N$ may contain
terms representing the dressing of the nucleon. (The
positive-energy projection operators may be omitted here, and
throughout the rest of this section, since they have no
effect, due to our use of the no-anti-nucleon approximation.)
Secondly, consider any time-ordered perturbation theory
Green's function:  \begin {equation} {\overline{G}_{NN}}_{(2)}
(E,\vec{p}_1^{\, \prime},\vec{p}_2^{\,
\prime},\vec{p}_1,\vec{p}_2) \end {equation} LSZ reduction
applied to ${\overline{G}_{NN}}_{(2)} (E,\vec{p}_1^{\,
\prime},\vec{p}_2^{\, \prime},\vec{p}_1,\vec{p}_2)$ implies:
\begin {eqnarray} {\overline{G}_{NN}}_{(2)} (E,&
&\vec{p}_1^{\, \prime},\vec{p}_2^{\,
\prime};\vec{p}_1,\vec{p}_2)=\nonumber\\ & &
\frac{i}{E^+-E_N(\vec{p}_1^{\, \prime})-E_N(\vec{p}_2^{\,
\prime})} {\overline{T}_{NN}}_{(2)} (E,\vec{p}_1^{\,
\prime},\vec{p}_2^{\, \prime},\vec{p}_1,\vec{p}_2)
\frac{i}{E^+-E_N(\vec{p}_1)-E_N(\vec{p}_2)}, \label
{eq:LSZTOPT} \end {eqnarray} where, once again, $E_N$ may
contain terms representing the dressing of the
nucleons.\footnote {Note that the two-nucleon propagator used
in this equation does not contain the full dressing on both
nucleons, since certain time-orders are excluded from it. As
discussed in Section {\protect \ref{sec-intro}}, this
inadequate dressing was the original reason for KB's
suggesting the use of a convolution integral for the free
two-nucleon Green's function {\protect \cite{BK92A}}.}  Above
we defined ${\overline{G}^\Sigma_{NN}}_{(2)}$ to be the sum of
all Green's functions differing only in the relative
time-order of processes which occur on different nucleons,
excluding transmitted pion emission and absorption.  Now we
also define:   \begin {equation}
{\overline{T}^\Sigma_{NN}}_{(2)}=\sum_{\mbox {All
time-orders}} {\overline{T}_{NN}}_{(2)}. \end {equation} Since
Eq.(\ref{eq:LSZTOPT}) holds between each term in this sum over
all time-orders, it must relate
${\overline{G}_{NN}^\Sigma}_{(2)}$ to
${\overline{T}_{NN}^\Sigma}_2$ as well. Therefore,
Eq.(\ref{eq:LSZTOPT}) for ${\overline{G}_{NN}^\Sigma}_{(2)}$
and Eq.(\ref{eq:LSZSPGF}) for ${G_{NN}}_{(1)}$ may be
substituted into Eq.(\ref{eq:BKnodelta}) in order to obtain
the relationship between ${\overline{T}_{NN}^{\Sigma}}_{(2)}$
and ${T_{NN}}_{(1)}$. When this is done we find:  \begin
{eqnarray} &{\overline{T}_{NN}^\Sigma}_{(2)}&(E,\vec{p}_1^{\,
\prime},\vec{p}_2^{\, \prime},\vec{p}_1,\vec{p}_2)=\int \frac
{dz \, dz'}{(-2 \pi i)^2} (\frac{1}{E^+-z'-E_N(\vec{p}_1^{\,
\prime})} +  \frac{1}{{z'}^+-E_N(\vec{p}_2^{\,
\prime})})\nonumber\\   \times
&{T_{NN}}_{(1)}&(E-z',\vec{p}_1^{\, \prime},z',\vec{p}_2^{\,
\prime};E-z,\vec{p}_1,z,\vec{p}_2)(\frac{1}{E^+-z-E_N(\vec{p}_1)}
+ \frac{1}{{z}^+-E_N(\vec{p}_2^{\, \prime})}) \label
{eq:TTreln} \end {eqnarray} \noindent If we wanted to
calculate a physical quantity from a sum of time-ordered
perturbation theory diagrams in which all relative time-orders
are included, we would need to calculate
${\overline{T}_{NN}^\Sigma}_{(2)}$ on-shell. However, it may
well be easier to calculate ${T_{NN}}_{(1)}$ and then use the
relation between ${T_{NN}}_{(1)}$ and
${\overline{T}_{NN}^\Sigma}_{(2)}$ to obtain
${\overline{T}^\Sigma_{NN}}_{(2)}$ on-shell.

In order to find out if this is feasible we set $E=E^{on}$ with
\begin {equation} E^{on}=E_N(\vec{p}_1) +
E_N(\vec{p}_2)=E_N(\vec{p}_1^{\, \prime}) + E_N(\vec{p}_2^{\,
\prime}). \end {equation} For this value of $E$
Eq.(\ref{eq:TTreln}) reads: \begin {eqnarray}
{T_{NN}^\Sigma}_{(2)} (E^{on},\vec{p}_1^{\,
\prime},\vec{p}_2^{\, \prime},\vec{p}_1,\vec{p}_2)&=&\int
\frac {dz dz'}{(-2 \pi i)^2} \mbox{disc}
(\frac{1}{z'-E_N(\vec{p}_2^{\, \prime})}) \nonumber\\ \times
{T_{NN}}_{(1)}(E^{on}-z',\vec{p}_1^{\,
\prime},z',\vec{p}_2^{\,
\prime};&E^{on}&-z,\vec{p}_1,z,\vec{p}_2)
\mbox{disc}(\frac{1}{z-E_N(\vec{p}_2)}).  \label {eq:OnTTreln}
\end {eqnarray} \noindent But: \begin {equation} \mbox
{disc}(\frac{1}{z-w})=-2 \pi i \delta(z-w). \end {equation}
Therefore, \begin {equation} {T_{NN}^\Sigma}_{(2)}
(E^{on},\vec{p}_1^{\, \prime},\vec{p}_2^{\,
\prime},\vec{p}_1,\vec{p}_2)= {T_{NN}}_{(1)}(E_N(\vec{p}_1^{\,
\prime}),\vec{p}_1^{\, \prime},E_N(\vec{p}_2^{\,
\prime}),\vec{p}_2^{\,
\prime};E_N(\vec{p}_1),\vec{p}_1,E_N(\vec{p}_2),\vec{p}_2).
\end {equation} This is a more general version of a result
obtained by Kvinikhidze and Blankleider by different means
\cite{BK92B}. It states that, on-shell, the amplitude
obtained  from the no-anti-nucleon two-nucleon Green's
function is equal to the sum over all time-orders of the
amplitudes obtained from time-ordered perturbation theory.

\section {The method of Klein, L\'evy, Macke and Kadyshevsky (KLMK)}

\label {sec-KLMK}

In the 1950s and 60s there was considerable interest in techniques
for reducing the Bethe-Salpeter (BS) equation to a
three-dimensional integral equation. In this section we explain
how this work is connected to our work above and in so doing
reveal that the KLMK method suffers from similar inconsistencies
to those of the KB method discussed above.

Initial work by Salpeter showed that if an instantaneous
interaction was used in the Bethe-Salpeter equation a
three-dimensional integral equation for the positive-energy
component of the wave function was obtained \cite{Sa52}. This work
was then extended by L\'evy and Klein, who showed that, in the
ladder approximation, the BS equation could be approximated by the
following integral equation for the wave function
$\varphi_{++}(\vec{p}^{\,})$ \cite{Le52A,Kl53,KM58}:  \begin
{equation} (E-2E_N(\vec{p}^{\,})) \varphi_{++}(\vec{p}^{\,})=
\Lambda_+^{(1)}(\vec{p}^{\,}) \Lambda_+^{(2)}(-\vec{p}^{\,})  \int
d^3k \, V_{OPE} (E,\vec{p},\vec{p}-\vec{k})
\varphi_{++}(\vec{p}-\vec{k}),  \end {equation} where the wave
function $\varphi_{++}(\vec{p}^{\,})$ is defined by: \begin
{eqnarray} (E- && 2E_N(\vec{p}^{\,}))
\varphi_{++}(\vec{p}^{\,})=\nonumber\\ && (-2 \pi i) (\frac{1}{2}
E + p_0 - E_N(\vec{p}^{\,})) (\frac{1}{2} E - p_0 -
E_N(-\vec{p}^{\,})) \Lambda_+^{(1)}(\vec{p}^{\,})
\Lambda_+^{(2)}(-\vec{p}^{\,}) \psi(p),  \end {eqnarray} with $p$
the relative four-momentum and $E$ the total energy in the
center-of-mass frame. The kernel
$V_{OPE}(E,\vec{p},\vec{p}-\vec{k})$ is given by:  \begin
{eqnarray} V_{OPE}(E,\vec{p},\vec{p}-\vec{k})=-\lambda
(E-2E_N(\vec{p}^{\,})) [(\frac{1}{2 \pi i})^2 \int d^3k \, dk_0 \,
dp_0 \frac{1}{\frac{1}{2} E+p_0-E_N(\vec{p}^{\,})} \nonumber\\
\times \frac{1}{\frac{1}{2} E-p_0-E_N(-\vec{p}^{\,})}
\Gamma^{(1)} (p,p-k) \frac{1}{k^2 + \mu^2} \Gamma^{(2)} (-p,-p+k)
\nonumber\\ \times \frac{1}{\frac{1}{2}
E+p_0-k_0-E_N(\vec{p}-\vec{k})} \frac{1}{\frac{1}{2}
E-p_0+k_0-E_N(\vec{k}-\vec{p}^{\,})}] (E-2E_N(\vec{p}-\vec{k})).
\label {eq:VOPE} \end {eqnarray} If we regard $E_N$ as containing
a self-energy term for the nucleon then this kernel contains both
fully dressed vertices and fully dressed propagators and we may
write: \begin {equation}
V_{OPE}(E,\vec{p},\vec{p}-\vec{k})=(E-2E_N(\vec{p}^{\,}))
{\overline{G}_{NN}^\Sigma}_{(2)}
(E,\vec{p},-\vec{p},\vec{p}-\vec{k},\vec{k}-\vec{p}^{\,})
(E-2E_N(\vec{p}-\vec{k})), \end {equation} where
${\overline{G}_{NN}^\Sigma}_{(2)}
(E,\vec{p},-\vec{p},\vec{p}-\vec{k},\vec{k}-\vec{p}^{\,})$ is the
sum of the convolution integral Green's functions for pion
exchange from $N1$ to $N2$ and $N2$ to $N1$, which are given
respectively, by Eq.(\ref{eq:CIdOPE}) as written, and by the same
equation with $N1$ and $N2$ swapped. Note that in order to
establish this fact we need to perform the following
substitutions, which merely correspond to changing the frame of
reference to the centre-of-mass frame for the two-nucleon
system:   \begin {eqnarray} \vec{p}=\vec{p}_1^{\,
\prime}=-\vec{p}_2^{\, \prime};\\
 p_0=\frac{1}{2} E -z';\\ \vec{k}=\vec{p}_2-\vec{p}_2^{\,
\prime}=\vec{p}_1^{\, \prime}-\vec{p}_1;\\ k_0=z-z'. \end
{eqnarray}

Using Eq.(\ref{eq:LSZTOPT}) we see that this $V_{OPE}$ is the
amplitude for the sum of all the time-ordered perturbation theory
diagrams which represent the different time-orders of
fully-dressed one-pion exchange.

Klein went on to generalize this result, showing how the BS
equation: \begin {equation} M=I + IGM, \label {eq:BSE} \end
{equation} where $M$ and $I$ are regarded as operators in the
Hilbert space of energy-momentum states $|p \rangle$ could always
be written as a three-dimensional equation of the form:  \begin
{equation} T(E)=V(E)+V(E)G_{NN}^{++}(E)T(E), \label {eq:D3IE} \end
{equation} where \begin {equation} G_{NN}^{++}(E)=\Lambda_+^{(1)}
\Lambda_+^{(2)} G_{NN}(E), \end {equation} with the t-matrix
$T(E)$ defined by: \begin {eqnarray} T(E)=(\frac{1}{2 \pi i})^2
G_{NN}^{-1}(E) [\int dp_0' dp_0 G_N^{(+)}(\frac{1}{2} E - p_0')
G_N^{(+)}(\frac{1}{2} E + p_0') \nonumber\\ \times M(p_0',p_0;E)
G_N^{(+)}(\frac{1}{2} E - p_0) G_N^{(+)}(\frac{1}{2} E + p_0)]
G_{NN}^{-1}(E), \end {eqnarray} and the potential $V(E)$ given by:
\begin {equation} V(E)=V_1(E) + V_2(E) - V_1(E) G^{++}_{NN}(E)
V_1(E) + \ldots, \label{eq:VE} \end {equation} where: \begin
{eqnarray} V_1(E)=(\frac{1}{2 \pi i})^2 G_{NN}^{-1}(E) [\int dp_0'
dp_0 G_N^{(+)}(\frac{1}{2} E - p_0') G_N^{(+)}(\frac{1}{2} E +
p_0') \nonumber\\ \times I(p_0',p_0;E) G_N^{(+)}(\frac{1}{2} E -
p_0) G_N^{(+)}(\frac{1}{2} E + p_0)] G_{NN}^{-1}(E),\\
V_2(E)=(\frac{1}{2 \pi i})^2 G_{NN}^{-1}(E) [\int dp_0' dk_0 dp_0
G_N^{(+)}(\frac{1}{2} E - p_0') G_N^{(+)}(\frac{1}{2} E + p_0')
I(p_0',k_0;E) \nonumber\\ \times G(k_0,E) I(k_0,p_0;E)
G_N^{(+)}(\frac{1}{2} E - p_0) G_N^{(+)}(\frac{1}{2} E + p_0)]
G_{NN}^{-1}(E) \ldots. \end {eqnarray} All operators in these
formulae are now considered to be operators in three-dimensional
momentum space.

Note that the propagator $G_{NN}(E)$ is the time-ordered
perturbation theory propagator: \begin {equation}
G_{NN}(E)=\frac{1}{E^+-E_N-E_N}. \label {eq:prop} \end {equation}
If dressed particles are involved the use of this time-ordered
perturbation theory propagator is open to question \`{a} la KB
\cite{BK92A}. But, we ignore this difficulty for the present and
assume that this $G_{NN}(E)$ is the correct propagator to use,
since any such problem may be fixed by making a minor modification
to the approach outlined here.

Now if the interaction kernel $I$ contains only the single pion
exchange that gives the ladder BS equation then:  \begin {equation}
\langle \vec{p}^{\, \prime} |V_1(E)|\vec{p}
\rangle=V_{OPE}(E,\vec{p}^{\, \prime},\vec{p}^{\,}). \end
{equation}

This result was also obtained by Kadyshevsky and his collaborators
\cite{Ka68,It70}, by proceeding from the Lippman-Schwinger
equation for the amplitude. Klein's early work focused on deriving
this result for the wave function BS equation, although he later
also derived it for the BS equation for amplitudes \cite{KL74}.

Klein also explained how to obtain a time-ordered perturbation
theory expression for $V(E)$ from the results given here
\cite{Kl54}. This explanation indicated that the potential $V(E)$
could be regarded as the sum of all possible time-ordered
perturbation theory diagrams for the process under consideration.
That is to say, given any BS equation interaction kernel $I$,
there are two equivalent ways of obtaining the three-dimensional
integral equation (\ref{eq:D3IE}) corresponding to the BS equation
(\ref{eq:BSE}):  \begin {enumerate} \item Use Eq.(\ref{eq:VE}) in
order to derive the equivalent energy-dependent potential $V(E)$;

\item Sum all two-particle irreducible time-ordered perturbation
theory diagrams allowed for this $I$; the result is $V(E)$.   \end
{enumerate} That these two methods are both equivalent to the BS
equation is a manifestation of the result derived above in Sections
\ref{sec-TwonucleonGF}--\ref{sec-BKconnection}: both the
convolution integral and time-ordered perturbation theory may be
obtained from field theory in the equal-time no-anti-nucleon
approximation.

Although in theory both of these two methods provide
three-dimensional integral equations containing exactly the same
information as the BS equation, in practice it is not feasible to
obtain the full $V(E)$ which is necessary in order to make the
equivalence exact. Consequently approximations to the potential,
such as that defined by Equation (\ref{eq:VOPE}), must be made.
Sometimes additional modifications to the propagator are also made
in order to further facilitate the calculation, see e.g.
\cite{It70}. However, here we merely wish to examine the quality
of the approximation: \begin {equation} V(E)=V_1(E),
\label{eq:VOPE2} \end {equation} which is obtained by taking the
first term in the series (\ref{eq:VE}) for $V$ when the BS
equation is being used in the ladder approximation, and so $I$
consists only of one-pion exchange. If dressed propagators are
being used the time-ordered perturbation theory diagrams which are
omitted from this approximation but included in the ladder BS
equation include those shown in Figure \ref{fig-KKomit}. These
omitted diagrams are to be contrasted with those in Figure
\ref{fig-KKinc}, which are included in the calculation using
Eqs.(\ref{eq:D3IE}) and (\ref{eq:VOPE2}). It is clear that, as is
the case in  the KB method and the current $NN-\pi NN$ equations,
different time-orders of the same physical process are treated
differently.

Connected to this problem is the problem of relative energy
non-conservation, which occurs here in a similar way to that in
which it occurred in the KB approximation discussed in Section
\ref{sec-KBapprox}. If we examine the energy assignments for the
two-pion exchange diagram generated by Equations (\ref{eq:D3IE})
and (\ref{eq:VOPE2}) we find that they correspond to those drawn
in Figure \ref{fig-TPE}. Although the energy assignments for
one-pion exchange are now correct (compare with Figures
\ref{fig-FullCIOPE} and \ref{fig-BKapOPE}) there is still a sudden
jump in energy where one one-pion exchange diagram, calculated
using a certain pair of relative energies, is joined to another
one-pion exchange diagram, which is calculated using different
relative energies. Again, as discussed above for the KB
approximation, this jump, could, in principle, be infinite, since
the relative energy variables $k_0$ and $k_0'$ are integrated
over. Consequently, an examination of KLMK's techniques shows that
although their work contains many of the same elements as those
pursued in this paper, ultimately it has a different purpose,
since it places a higher value on obtaining a tractable
three-dimensional integral equation than on including all possible
time-orders of the relevant processes. In this sense it is more
akin to the work of KB than to the approach espoused here, since
both KB and KLMK restrict themselves to considering only
three-dimensional integral equations and using energy-integrations
to sum certain pieces in their essentially three-dimensional
theories. They pay for this restriction however, since both
approaches violate relative energy conservation and neither
includes all time-orders of physically relevant processes. If
these two problems are to be eliminated a fully four-dimensional
calculation must be pursued.

\section {Conclusion}

\label {sec-conc}

Current models of the $NN-\pi NN$ system are unsatisfactory in two
ways. Firstly, they fail to correctly predict the experimental
data. Secondly, because all calculations are done, in some sense or
another, in a time-ordered framework, diagrams which are merely
different time-orders of the same process are treated
unequally---the most famous example of this being the
non-inclusion of the Jennings diagram in the current $NN-\pi NN$
equations \cite{Je88,JR88}. It is to be hoped that a resolution of
the second problem will, by producing a theoretically consistent
model, lead to a diminution of the first problem. Therefore, the
question is how to consistently include all time-orders in a
calculation.

In this paper we displayed a scheme for doing just this.
Starting from a Lagrangian closely related to those of Quantum
Hadrodynamics \cite{Wa74,Se92} we explained how to calculate fully
covariant one-nucleon, one-pion, two-nucleon and $m \rightarrow
m'$-particle Green's functions, either with or without
anti-nucleonic degrees of freedom. It was then shown that, in the
absence of anti-nucleonic degrees of freedom, {\em any} covariant
perturbation theory diagram may be expressed as the sum of a set
of time-ordered perturbation theory diagrams, provided we restrict
all particles to have the same time in the initial and final
states. This result made it  clear that Kvinikhidze and
Blankleider's original convolution integral formula for
disconnected diagrams \cite{BK92A} is merely a manifestation of
the more general relationship between covariant perturbation
theory and time-ordered perturbation theory. This fact allowed us
to point out the inconsistency in the KB approach to the $NN-\pi
NN$ problem \cite{BK92B}, which leads to a  violation of relative
energy conservation. To complete our derivation of the properties
of convolution integral theory we explained how LSZ reduction
\cite{LSZ55} could be performed on convolution integral Green's
functions. We then observed that KLMK's work
\cite{Le52A,Le52B,Kl53,Kl54,Kl58,KM58,Ma53A,Ma53B,KL74,Ka68,It70},
which also used a convolution integral to calulcate the one-pion
exchange potential, has problems similar to those which occur in
the KB approximation. These problems occur because any approximate
three-dimensional integral equation, which involves an energy
integration for the propagators, as in the work of KB, or for
one-pion exchange, as in the work of KLMK, results in relative
energy non-conservation. This is connected to the fact that such
three-dimensional integral equations always omit certain
time-orders of some physical process for which other time-orders
are included. And, if there is one thing the failure of the
current $NN-\pi NN$ equations teaches us, it is that we treat
different time-orders of the same physical process differently at
our peril. As stressed above, the only way to circumvent this
difficulty and consistently include {\em all} time-orders of the
relevant physical processes in a natural way is to use
time-dependent perturbation theory and derive four-dimensional
integral equations.

Having developed these points in this work the next task is the
application of the perturbation theory developed in the first five
sections of this paper in order to derive fully covariant
scattering equations for systems of nucleons and pions. The
technique chosen to do this is the classification-of-diagrams
method of Taylor~\cite{Ta63}. Because there are some subtleties in
the application of Taylor's method to perturbation schemes other
than time-ordered perturbation theory we will explain Taylor's
technique fully and elucidate these subtleties in a forthcoming
paper \cite{AP93B}. In a third paper the results developed in this
second paper will be applied to the $NN-\pi NN$ system
\cite{AP93C}.

\acknowledgements {We wish to acknowledge many interesting
discussions with  A.~N.~Kvinikhidze and B.~Blankleider, who first
proposed the use of a convolution integral in order to sum all
time-orders in disconnected time-ordered perturbation theory
diagrams. We are also grateful to T.-S.~H.~Lee who pointed out the
work of KLMK to us. D.~R.~P. wishes to acknowledge discussions with
C.~H.~M.~van Antwerpen and S.~B.~Carr. We both wish to thank the
Australian Research Council for their financial support. D.~R.~P.
holds an Australian Postgraduate Research Award.}

\appendix

\section {Calculation of the one-particle free Green's functions}

\label {ap-GFcalc}

In this appendix we calculate the one-nucleon and one-pion free
Green's functions. We then apply the no-anti-nucleon approximation to
these Green's functions, and hence obtain the form of the free
Green's functions in the no-anti-nucleon approximation.

\subsection {Calculation of the full covariant Green's functions}

We begin by noting that the Green's functions in question are defined
by: \begin {eqnarray} G_N^{(0)}(x',x)&=&\langle 0|T({\psi}^I(x')
{\overline{\psi}}^I(x))|0 \rangle, \label{eq:gn02}\\
{G_\pi}^{(0)}_{ji}(x',x)&=&\langle 0|T(\phi_j^I(x') \phi_i^I(x))|0
\rangle.  \label{eq:gpi02}  \end {eqnarray} Firstly, consider the
one-nucleon Green's function. Using the definition of the
time-ordering operator $T$, the expansions of the Schr\"odinger
representation field operators
$\psi(\vec{x}),\overline{\psi}(\vec{x})$: \begin {eqnarray}
\psi(\vec{x})&=&\int \frac {d^3k}{(2 \pi)^3} \frac{m}{E_N(\vec{k})}
\sum_{\alpha} [b_\alpha(\vec{k}) u_\alpha(\vec{k}) e^{i \vec{k} \cdot
\vec{x}} + d^{\dagger}_\alpha(\vec{k}) v_\alpha(\vec{k}) e^{-i
\vec{k} \cdot \vec{x}}], \label{eq:psi2}\\
\overline{\psi}(\vec{x})&=&\int \frac {d^3k}{(2 \pi)^3}
\frac{m}{E_N(\vec{k})} \sum_{\alpha} [b^{\dagger}_\alpha(\vec{k})
\overline{u}_\alpha(\vec{k}) e^{-i \vec{k} \cdot \vec{x}} +
d_\alpha(\vec{k}) \overline{v}_\alpha(\vec{k}) e^{i \vec{k} \cdot
\vec{x}}], \label {eq:psib2} \end {eqnarray} and the connection
between the Schr\"odinger and Interaction representations gives:
\begin {equation} {G_N^{(0)}}(x',x)={G_N^{(0)}}^+(x',x)
\theta(x_0'-x_0) + {G_N^{(0)}}^-(x',x) \theta(x_0-x_0'); \end
{equation} with: \begin {eqnarray} {G_N^{(0)}}^+(x',x)&=& \int \frac
{d^3p \, d^3p'}{(2 \pi)^6} \frac {m^2 \sum_{\alpha \alpha'}
u_{\alpha'}(\vec{p}^{\, \prime}) \overline{u}_\alpha(\vec{p}^{\,})}
{E_N(\vec{p}^{\,}) E_N(\vec{p}^{\, \prime})}  \langle
0|b_{\alpha'}(\vec{p}^{\, \prime}) e^{-i H_K (x'_0 - x_0)}
b^{\dagger}_\alpha (\vec{p}^{\,})|0 \rangle  e^{i(\vec{p}^{\, \prime}
\cdot \vec{x}^{\, \prime} - \vec{p} \cdot \vec{x})},\\
{G_N^{(0)}}^-(x',x)&=&\int \frac {d^3p \, d^3p'}{(2 \pi)^6} \frac
{m^2 \sum_{\alpha \alpha'}  v_{\alpha}(\vec{p}^{\,})
\overline{v}_{\alpha'}(\vec{p}^{\, \prime})}{ E_N(\vec{p}^{\,})
E_N(\vec{p}^{\, \prime})} \langle 0|d_\alpha(\vec{p}^{\,}) e^{-i H_K
(x_0 - x'_0)} d^{\dagger}_{\alpha'}(\vec{p}^{\, \prime})|0 \rangle
e^{i(\vec{p} \cdot \vec{x} - \vec{p}^{\, \prime} \cdot \vec{x}^{\,
\prime})}, \label{eq:GN-}  \end {eqnarray}   where we have simplified
${G_N^{(0)}}^+$ and ${G_N^{(0)}}^-$ by eliminating those terms which
are zero due to:  \begin {eqnarray} b_\alpha(\vec{p}^{\,})|0 \rangle
&=& 0,\\ d_\alpha(\vec{p}^{\,})|0 \rangle &=& 0. \end {eqnarray}

Now consider ${G_N^{(0)}}(p)$, defined by: \begin {equation}
{G_N^{(0)}}(p',p)=(2 \pi)^4 \delta^{(4)}(p-p'){G_N^{(0)}}(p) \end
{equation} where: \begin {equation} {G_N^{(0)}}(p',p)=\int
d^4x\,d^4x' e^{ip'x'} {G_N^{(0)}}(x',x)  e^{-ipx}={G_N^{(0)}}^+(p',p)
+ {G_N^{(0)}}^-(p',p). \end {equation} Here ${G_N^{(0)}}^+(p',p)$ and
${G_N^{(0)}}^-(p',p)$ are the Fourier transforms of
${G_N^{(0)}}^+(x',x) \theta(x_0'-x_0)$ and ${G_N^{(0)}}^-(x',x)
\theta(x_0-x_0')$ respectively.  Evaluation of  ${G_N^{(0)}}^+(p',p)$
gives:  \begin {eqnarray} {G_N^{(0)}}^+(p',p)&=&\int dx_0 \, dx_0'
\frac {m^2\sum_{\alpha,\alpha'} u_{\alpha'} (\vec{p}^{\, \prime})
\overline{u}_{\alpha}(\vec{p}^{\,})}{E_N(\vec{p}^{\,})
E_N(\vec{p}^{\, \prime})}\nonumber\\ & &\langle
0|b_{\alpha'}(\vec{p}^{\, \prime}) e^{-i H_K (x_0' - x_0)}
b^{\dagger}_\alpha(\vec{p}^{\,})|0 \rangle  \theta(x_0'-x_0)
e^{i(p_0'x_0' - p_0 x_0)}. \end {eqnarray} Transforming to relative
and average times: \begin {equation} T=\frac{x_0 + x_0'}{2},
\tau=x_0' - x_0; \end {equation} and noting that $dx_0 dx_0'=dT
d\tau$ gives: \begin {eqnarray} &{G_N^{(0)}}^+& (p',p)\nonumber\\ &=&
\int dT e^{i (p_0' - p_0) T} \int_0^\infty d\tau \, \frac {m^2
\sum_{\alpha,\alpha'} u_{\alpha'} (\vec{p}^{\, \prime})
\overline{u}_{\alpha} (\vec{p}^{\,})}{E_N(\vec{p}^{\,})
E_N(\vec{p}^{\, \prime})} \langle 0|a_{\alpha'}(\vec{p}^{\, \prime})
e^{i [(\frac{p_0' + p_0}{2}) - H_K] \tau}
a^{\dagger}_\alpha(\vec{p}^{\,})|0 \rangle \\  &=&2 \pi \delta
(p_0'-p_0) \frac{m^2 \sum_{\alpha \alpha'} u_{\alpha'}(\vec{p}^{\,
\prime}) \overline{u}_\alpha (\vec{p}^{\,})}{E_N(\vec{p}^{\,})
E_N(\vec{p}^{\, \prime})}  \lim_{\epsilon \rightarrow 0+}
\int_0^\infty d\tau \langle 0|a_{\alpha'}(\vec{p}^{\, \prime})
e^{i(p_0 + i\epsilon - H_K) \tau} a^{\dagger}_\alpha(\vec{p}^{\,})|0
\rangle \\  &=&2 \pi \delta (p_0'-p_0) \frac{m^2}{E_N(\vec{p}^{\,})
E_N(\vec{p}^{\, \prime})} \sum_{\alpha \alpha'}
u_{\alpha'}(\vec{p}^{\, \prime})  \overline{u}_\alpha(\vec{p}^{\,})
\lim_{\epsilon \rightarrow 0+} \frac{i}{p_0 + i\epsilon -
E_N(\vec{p}^{\, \prime})}\langle \vec{p}^{\, \prime} \alpha'|\vec{p}
\, \alpha \rangle,  \end {eqnarray}  where $H_K|\vec{p}^{\, \prime}
\alpha' \rangle =E_N(\vec{p}^{\, \prime})|\vec{p}^{\, \prime} \alpha'
\rangle$.  Now the Fock space states $|\vec{p} \, \alpha \rangle $
are normalized such that:  \begin {equation} \langle \vec{p}^{\,
\prime} \alpha'|\vec{p} \, \alpha \rangle =\delta_{\alpha \alpha'} (2
\pi)^3 \frac{E_N(\vec{p}^{\,})}{m} \delta^{(3)}(p'-p), \end
{equation} hence: \begin {equation} {G_N^{(0)}}^+(p',p)=(2 \pi)^4
\delta^{(4)}(p'-p) \frac {im}{E_N(\vec{p}^{\,})} \frac {\sum_\alpha
u_\alpha(\vec{p}^{\,}) \overline{u}_\alpha(\vec{p}^{\,})}{p_0^+ -
E_N(\vec{p}^{\,})}. \end {equation} Therefore, \begin {equation}
{G_N^{(0)}}^+(p)=\frac{im}{E_N(\vec{p}^{\,})}\frac {\sum_\alpha
u_\alpha(\vec{p}^{\,}) \overline{u}_\alpha(\vec{p}^{\,})}{p_0^+ -
E_N(\vec{p}^{\,})}, \label {eq:GN0+} \end {equation} in agreement
with Garcilazo  up to a factor of $i$ \cite{Ga93}. This result is
also in agreement with van Faassen, again up to a factor of $i$. van
Faassen absorbs the factor of $\frac{m}{E_N(\vec{p}^{\,})}$ into his
spinor normalization \cite{vFa85}.

A similar calculation for ${G_N^{(0)}}^-(p)$ yields: \begin {equation}
{G_N^{(0)}}^-(p)=\frac {im}{E_N(-\vec{p}^{\,})} \frac {\sum_\alpha
v_\alpha(-\vec{p}^{\,}) \overline{v}_\alpha(-\vec{p}^{\,})}{p_0^- +
E_N(-\vec{p}^{\,})}, \end {equation} again in agreement with
Garcilazo. Noting that $E_N(-\vec{p}^{\,})=E_N(\vec{p}^{\,})$ these
two results may be combined  in order to produce:  \begin {equation}
{G_N^{(0)}}(p)=\frac{im}{E_N(\vec{p}^{\,})} \left[ \frac{\sum_\alpha
u_\alpha(\vec{p}^{\,}) \overline{u}_\alpha(\vec{p}^{\,})}{p_0^+ -
E_N(\vec{p}^{\,})} + \frac{\sum_\alpha v_\alpha(-\vec{p}^{\,})
\overline{v}_\alpha(-\vec{p}^{\,})}{p_0^- + E_N(\vec{p}^{\,})}
\right],  \end {equation} which may then be simplified to yield:
\begin {equation} G_N^{(0)}(p)=\frac{i}{\not\!{p} - m}, \end
{equation} in accordance with Eq.(\ref{eq:GNcov}).

We now turn to the free pion Green's function: \begin {equation}
{G_\pi}^{(0)}_{ji}(x',x)=\langle 0|T(\phi_j^I(x') \phi_i^I(x))|0
\rangle . \end {equation} Using the expansion of the Schr\"odinger
operator: \begin {equation} \phi_i(\vec{x})=\int \frac{d^3k}{(2
\pi)^3 2 \omega_\pi(\vec{k})} [{a_i}(\vec{k}) e^{i\vec{k} \cdot
\vec{x}} + {a^\dagger_i} (\vec{k}) e^{-i \vec{k} \cdot \vec{x}}] \end
{equation} and the relation  of the interaction and Schr\"odinger
operators gives, by a similar argument to that for the free
one-nucleon Green's function:  \begin {equation}
{{G_\pi}^{(0)}_{ji}}(x',x)={{G_\pi}^{(0)}_{ji}}^+(x',x)
\theta(x_0'-x_0) + {{G_\pi}^{(0)}_{ji}}^-(x',x) \theta(x_0-x_0'),
\end {equation} with: \begin {eqnarray}
{{G_\pi}^{(0)}_{ji}}^+(x',x)&=&\int \frac {d^3k \, d^3k'}{(2 \pi)^6}
\frac{1}{2 \omega_\pi(\vec{k}) 2 \omega_\pi(\vec{k}')} \langle
0|a_j(\vec{k}') e^{-i H_K (x_0' - x_0)} a^{\dagger}_i(\vec{k})|0
\rangle  e^{i(\vec{k}' \cdot \vec{x}^{\, \prime} - \vec{k} \cdot
\vec{x})},\\  {{G_\pi}^{(0)}_{ji}}^-(x',x)&=&\int \frac {d^3k \,
d^3k'}{(2 \pi)^6} \frac {1}{2 \omega_\pi(\vec{k}) 2
\omega_\pi(\vec{k}')} \langle 0|a_i(\vec{k}) e^{-i H_K (x_0 - x_0')}
a^{\dagger}_j(\vec{k}')|0 \rangle   e^{i(\vec{k} \cdot \vec{x} -
\vec{k}' \cdot \vec{x}^{\, \prime})}. \end {eqnarray} If we define
${{G_\pi}^{(0)}_{ji}}^+(k',k)$ and ${{G_\pi}^{(0)}_{ji}}^-(k',k)$ to
be the Fourier transforms of ${{G_\pi}^{(0)}_{ji}}^+(x',x)
\theta(x_0'-x_0)$ and ${{G_\pi}^{(0)}_{ji}}^-(x',x)
\theta(x_0-x_0')$, we may obtain, by a similar procedure to that used
for the free one-nucleon Green's function above:    \begin {eqnarray}
{{G_\pi}^{(0)}_{ji}}^+(k',k)&=&(2 \pi)^4 \delta^{(4)}(k'-k) \frac{i}{2
\omega_\pi(\vec{k})} \, \frac{\delta_{ij}}{k_0^+ -
\omega_\pi(\vec{k})},\\ {{G_\pi}^{(0)}_{ji}}^-(k',k)=&-&(2 \pi)^4
\delta^{(4)}(k'-k) \frac{i}{2 \omega_\pi(-\vec{k})} \,
\frac{\delta_{ij}}{k_0^- + \omega_\pi(-\vec{k})}. \end {eqnarray}

Note that once again, these formulae agree with Garcilazo up to a
factor of $i$ \cite{Ga93}. Note also that: \begin {equation}
{{G_\pi}^{(0)}_{ji}}^-(k',k)={{G_\pi}^{(0)}_{ji}}^+(-k,-k'). \end
{equation}

If we use the fact $\omega_\pi(\vec{k})=\omega_\pi(-\vec{k})$, these
may be combined to give the total Green's function:  \begin {equation}
{G_\pi}^{(0)}_{ji}(k',k)=(2 \pi)^4 \delta^{(4)}(k'-k)
{G_\pi}^{(0)}_{ji}(k). \end {equation} with: \begin {equation}
{G_\pi}^{(0)}_{ji}(k)=\frac{i \delta_{ij}}{k^2 - m_\pi^2}, \end
{equation} in agreement with Eq.(\ref{eq:Gpicov}).

\subsection {The no-anti-nucleon approximation}

So far we have not made any approximations in the calculation of
these Green's functions. In this section we examine what happens if
we make the no-anti-nucleon approximation, i.e.  take the limit
$v(\vec{p}^{\,}) \rightarrow 0$ and $\overline{v}(\vec{p}^{\,})
\rightarrow 0$. Examining Eq.(\ref{eq:GN-}) reveals, in this limit,
we obtain:  \begin {equation} {G_N^{(0)}}^-(x',x)=0, \end {equation}
and so, \begin {equation} G_N^{(0)}(x',x)={G_N^{(0)}}^+(x',x)
\theta(x_0'-x_0), \end {equation} in the no-anti-nucleon
approximation. This proves Eq.(\ref{eq:thetaG}).

Furthermore, since: \begin {equation}
G_N^{(0)}(p',p)={G_N^{(0)}}^+(p',p) + {G_N^{(0)}}^-(p',p), \end
{equation} and ${G_N^{(0)}}^-(p,p')$ is zero in this approximation,
it follows that: \begin {equation}
G_N^{(0)}(p',p)={G_N^{(0)}}^+(p',p). \end {equation} Consequently,
Eq.(\ref{eq:GN0+}) implies that, in the no-anti-nucleon approximation:
\begin {equation}  G_N^{(0)}(p)=\frac{im}{E_N(\vec{p}^{\,})}\frac
{\sum_\alpha u_\alpha(\vec{p}^{\,})
\overline{u}_\alpha(\vec{p}^{\,})}{p_0^+ - E_N(\vec{p}^{\,})}, \end
{equation} as claimed in Eq.(\ref{eq:pfreeNGF}).

\section {Rewriting the perturbation expansion of the two-nucleon
Green's function in order to obtain old-fashioned perturbation theory}

\label {ap-changeonetotwo}

In this Appendix we establish that the vertex containing
single-particle operators, i.e.: \begin {equation} {\cal
H}_{int}(x)=ig \int d^4x_N \, d^4x_N' \, d^4x_\pi \overline{\psi}
(x_N') \gamma_5 \vec{\tau} \psi(x_N) \cdot  \vec{\phi}(x_\pi)
\Gamma(x-x_N',x-x_N,x-x_\pi), \end {equation} may be re-expressed in
terms of two-nucleon operators, which in the Schr\"odinger
representation are defined by:  \begin {eqnarray}
\psi_{NN}(\vec{x}_1,\vec{x}_2)&=&\psi(\vec{x}_1) \psi (\vec{x}_2)
\label{eq:psiNN},\\
\overline{\psi}_{NN}(\vec{x}_1,\vec{x}_2)&=&\overline{\psi}(\vec{x}_2)
\overline{\psi} (\vec{x}_1). \label {eq:bpsiNN}  \end {eqnarray} It
immediately follows that the equal-time two-nucleon Green's function
may be rewritten in terms of these two-particle operators without
changing its value, as was claimed in Section \ref{sec-BKconnection}.

{\bf Claim:} {\em Suppose that the state $|S \rangle ^I$ contains two
nucleons, and both these nucleons in $|S \rangle ^I$ are created at
the same time. Define ${{\cal H}^I_{int}}_{(2)}$ by:   \begin
{eqnarray} {{\cal H}^I_{int}}_{(2)} (y_0,\vec{y},\vec{y}^{\,
\prime})&=&\overline{\psi}^I (y_0,\vec{y}^{\, \prime}) {\cal
H}^I_{int}(y) \psi^I (y_0,\vec{y}^{\, \prime})\\ &=& ig \int d^4y_N
d^4y_N' d^4y_\pi \overline{\psi}^I_{NN} ({y_N'}_0,\vec{y}_N^{\,
\prime},\vec{y}^{\, \prime}) \gamma_5 \vec{\tau} \nonumber\\  & &
\times \psi_{NN}^I ({y_N}_0,\vec{y}_N,\vec{y}^{\, \prime}) \cdot
\vec{\phi}^I (y_\pi) \Gamma(y-y_N',y-y_N,y-y_\pi).  \end {eqnarray}
Then: \begin {equation} {\cal H}^I_{int}(y)|S \rangle ^I=\int d^3y'
{{\cal H}_{int}}_{(2)} (y_0,\vec{y},\vec{y}^{\, \prime})|S \rangle ^I.
\label {eq:claim} \end {equation} (Naturally these statements, if
true in one representation, are true in any representation. We have
chosen the interaction representation, because, as will be seen in
Eq.(\ref{eq:GNNresult}), it is operators in that representation which
the Green's function is expressed in terms of.)}

{\bf Proof:}

Consider: \begin {equation} \int d^3y' {{\cal H}_{int}}_{(2)}
(y_0,\vec{y},\vec{y}^{\, \prime})|S \rangle ^I, \end {equation} where
$|S \rangle ^I$ is a state in the interaction representation
containing an, as yet,  unspecified number of nucleons, all of which
were created at the same time. Firstly we note that, using the
anti-commutation relations of the $\overline{\psi}$s and $\psi$s, and
the definition of ${\cal H}^I_{int}$, we may obtain: \begin
{equation} {{\cal H}_{int}}_{(2)} (y_0,\vec{y},\vec{y}^{\,
\prime})={\cal H}^I_{int}  (y_0,\vec{y}^{\, })  e^{i H_K y_0}
\overline{\psi}(\vec{y}^{\, \prime}) \psi(\vec{y}^{\, \prime}) e^{-i
H_K y_0} - {\cal H}^I_{int} (y_0,\vec{y}^{\, }) \delta^{(3)}(y-y').
\end {equation}  Now if the state $|S \rangle ^I$ consists of
nucleons which are all created at the same time $t$ and pions created
at unspecified times it may be written, in the interaction
representation, as:    \begin {equation} |S \rangle ^I=
\prod_{k=1}^{n^\pi_S} {\phi^{\dagger}}^I(z_k) \prod_{j=1}^{n^N_S}
e^{i H_K t} \overline{\psi}(\vec{x}_j)|0 \rangle,  \end {equation}
where $n^N_S$ and $n^\pi_S$ are the numbers of nucleons and pions in
the state $|S \rangle $ and
$(t,\vec{x}_1),\ldots,(t,\vec{x}_{n^N_S})$ and
$z_1,\ldots,z_{n^\pi_S}$ are the space-time points at which these
nucleons and pions are created. Clearly, the operator $(\int d^3y'
\overline{\psi}(\vec{y}^{\, \prime}) \psi(\vec{y}^{\, \prime}))$
commutes with all the pionic operators, and, it may also be shown to
commute with $H_K$, leaving us with:   \begin {eqnarray} \int d^3y' &
& {{\cal H}_{int}}_{(2)} (y_0,\vec{y},\vec{y}^{\, \prime})|S \rangle
=\nonumber\\ & &{\cal H}^I_{int} (y_0,\vec{y}^{\, })
\prod_{k=1}^{n^\pi_S} {\phi^{\dagger}}^I(z_k) e^{i H_K t} (\int d^3y'
\overline{\psi}(\vec{y}^{\, \prime}) \psi(\vec{y}^{\, \prime}))
\prod_{j=1}^{n^N_S} \overline{\psi}(\vec{x}_j)|0 \rangle  - {\cal
H}^I_{int} (y_0,\vec{y}^{\, })|S \rangle .  \label {eq:Hchangefirst}
\end {eqnarray} Furthermore, from the anti-commutation relations of
the $\psi$s and $\overline{\psi}$s we find that: \begin {equation}
(\int d^3y' \overline{\psi}(\vec{y}^{\, \prime}) \psi(\vec{y}^{\,
\prime})) \overline{\psi}(\vec{x})=\overline{\psi}(\vec{x}) +
\overline{\psi}(\vec{x}) (\int d^3y' \overline{\psi}(\vec{y}^{\,
\prime}) \psi(\vec{y}^{\, \prime})). \end {equation} This implies
that: \begin {equation} (\int d^3y' \overline{\psi}(\vec{y}^{\,
\prime}) \psi(\vec{y}^{\, \prime})) \prod_{j=1}^{n^N_S}
\overline{\psi}(\vec{x}_j)=n^N_S \prod_{j=1}^{n^N_S}
\overline{\psi}(\vec{x}_j) + \prod_{j=1}^{n^N_S}
\overline{\psi}(\vec{x}_j) (\int d^3y' \overline{\psi}(\vec{y}^{\,
\prime}) \psi(\vec{y}^{\, \prime})), \end {equation} suggesting that
$(\int d^3y' \overline{\psi}(\vec{y}^{\, \prime}) \psi(\vec{y}^{\,
\prime}))$ may be interpreted as the fermionic number operator, as we
would expect. Substituting this in Eq.(\ref{eq:Hchangefirst}) and
simplifying the result gives:  \begin {equation} \int d^3y' {{\cal
H}^I_{int}}_{(2)} (y_0,\vec{y},\vec{y}^{\, \prime})|S \rangle
^I=(n^N_S-1){\cal H}^I_{int} (y_0,\vec{y}^{\, })|S \rangle ^I, \end
{equation} which, if $n^N_S=2$ proves Eq.(\ref{eq:claim}), thus
proving the claim.

Now consider the definition of $G_{NN}(x_1',x_2';x_1,x_2)$,
Eq.(\ref{eq:GNNdef}). By methods covered in most text books on field
theory, the following expression for $G_{NN}(x_1,x_2;x_1',x_2')$ may
be derived:   \begin {eqnarray}
G_{NN}(x_1',x_2';&x_1&,x_2)=\sum_{n=0}^{\infty} \frac {(-i)^n}{n!}
\int d^4y_1 d^4y_2 \cdots d^4y_n \nonumber\\ \times \langle
0|T(\psi^I(&x_1'&) \psi^I(x_2') {\cal H}^I_{int} (y_n) \ldots {\cal
H}^I_{int} (y_2) {\cal H}^I_{int} (y_1) {\overline{\psi}}^I(x_1)
{\overline{\psi}}^I(x_2))|0 \rangle _{con}, \label {eq:GNNresult}
\end {eqnarray} where the $con$ indicates that those contributions to
the matrix element which contain vacuum-vacuum subdiagrams are to be
ignored. Suppose that we make  the no-anti-nucleon approximation.
Suppose that we also consider only the equal-time Green's function,
i.e. we set $x_1^0=x_2^0 \equiv t$ and ${x_1^0}'={x_2^0}' \equiv t'$.
We begin by using the definition of $T$ in order to rewrite this
Green's function as a sum over all possible time-orders of the field
operators in the vacuum expectation value. Each time-order is written
as a separate term in the sum, with the integration over $y_1^0$,
$y_2^0$, \ldots, $y_n^0$ being suitably restricted.

We wish to prove that, if we consider any term in the sum over all
time-orders, replacing ${\cal H}_{int}(y)$ by $\int d^3y' {{\cal
H}_{int}}_{(2)} (y_0,\vec{y},\vec{y}^{\, \prime})$ does not change the
value of that particular term. So consider any one term in the sum
over all time-orders. Firstly note that if the presence of
anti-nucleons is forbidden then the time $t$ must be the earliest
time, and consequently the operators ${\overline{\psi}}^I(x_1)$ and
${\overline{\psi}}^I(x_2)$ must be the first to act. Now the state
${\overline{\psi}}^I(x_1) {\overline{\psi}}^I(x_2)|0 \rangle $ has a
nucleon number of two, and since nucleon number is conserved in this
theory, it follows that the number of nucleons in the state on which
${\cal H}_{int}$ is acting is always two. Furthermore, if all
vertices ${\cal H}_{int}(y)$ up to the $j$th one have been replaced
by vertices $\int d^3y' {{\cal H}_{int}}_{(2)} (y_0,\vec{y},\vec{y}^{\,
\prime})$ then it is clear that the $j$th interaction Hamiltonian
${\cal H}_{int}(y_j)$ acts on a state in which two nucleons, created
at the same time, are present. Consequently, ${\cal H}_{int}(y_j)$
may also be replaced by a vertex $\int d^3y_j' {{\cal H}_{int}}_{(2)}
(y_j^0,\vec{y}_j,\vec{y}_j^{\, \prime})$. Since the first ${\cal
H}_{int}(y)$ acts directly on the state ${\overline{\psi}}^I(x_1)
{\overline{\psi}}^I(x_2)|0 \rangle $ which clearly contains two
nucleons, which, by assumption, are created at the same time, it
follows, by induction, that replacing all vertices ${\cal
H}_{int}(y)$ by vertices $\int d^3y' {{\cal H}_{int}}_{(2)}
(y_0,\vec{y},\vec{y}^{\, \prime})$ does not change the value of this
term in the Green's function. Consequently, Eq.(\ref{eq:GNNresult})
may be rewritten:    \begin {eqnarray}
G_{NN}(x_1',x_2';x_1,x_2)=\sum_{n=0}^{\infty} \frac {(-i)^n}{n!}
\sum_{\mbox{All TOs}} \int_{\Omega} dy_1^0 d^3y_1 d^3&y_1'& dy_2^0
d^3y_2 d^3y_2'\ldots dy_n^0 d^3y_n d^3y_n'\nonumber\\ \times \langle
0|\psi^I_{NN}(t,\vec{x}_1^{\, \prime},\vec{x}_2^{\, \prime})  {{\cal
H}^I_{int}}_{(2)}  (y_n^0,\vec{y}_n,\vec{y}_n^{\, \prime}) \ldots
{{\cal H}^I_{int}}_{(2)} (y_1^0,&\vec{y}_1&,\vec{y}_1^{\, \prime})
\overline{\psi}^I_{NN}(x_1^0,\vec{x}_1,\vec{x}_2)|0 \rangle _{con},
\label{eq:GNNsum} \end {eqnarray}  where $\Omega$ is a region of
integration appropriate for the particular time-order under
consideration. Thus, we have proved that, in the absence of
anti-nucleons, the equal-time two-nucleon Green's function may be
rewritten in terms of the two-particle Green's functions without
changing its value.

\begin {figure} \vspace{33 mm} \hskip 54 mm \caption {The
so-called Jennings term, which is not included in the $NN-\pi
NN$ equations.} \label {fig-Jennings} \end {figure}

\begin {figure} \vspace {18 mm} \hskip 29.5 mm \vskip 3 mm
\caption {The diagram on the left is not included in the $NN$
propagator used in the $NN-\pi NN$ equations, while the
diagram on the right is.} \label {fig-NNdress} \end {figure}

\begin {figure}  \vspace {35 mm} \hskip 55 mm \vskip 3 mm
\caption {This diagram is currently included in the $NN-\pi
NN$ equations, even though it is merely a different time-order
of the excluded Jennings term.}  \label {fig-Included}  \end
{figure}

\begin {figure} \vspace {28 mm} \hskip 29 mm \vskip 3 mm
\caption {The diagram on the right is not included in the
Stingl and Stelbovics model, since it involves a three-pion
intermediate state. This occurs even though it is merely a
different time order of the left-hand diagram, which is
included in the model.} \label {fig-SSprob} \end {figure}

\vfill \eject \null

\begin {figure} \vspace{18 mm} \hskip 38 mm \vskip 3 mm
\caption {The $\pi NN$ vertex we will use in this work.}
\label {fig-vertex} \end {figure}

\begin {figure} \vspace{25 mm} \hskip 31 mm \vskip 3 mm
\caption {The diagram on the left contains a vacuum-vacuum
subdiagram and consequently is not included in the
perturbation expansion for the two-nucleon Green's function.
However, the diagram on the right, although it is disconnected
like the left-hand diagram, contains no vacuum-vacuum
subdiagram and consequently is included in the perturbation
expansion.}   \label {fig-vacvac}   \end {figure}

\begin {figure} \vspace{43 mm} \hskip 38 mm \vskip 3 mm
\caption {A diagrammatic representation of the two-nucleon
vertex, in which one nucleon is a spectator. To be compared to
the vertex in Figure  \protect {\ref{fig-vertex}} which is
only a one-nucleon vertex.}   \label{fig-newvertex}  \end
{figure}

\begin {figure} \vspace{20 mm} \hskip 4 mm \vskip 3 mm
\caption {A diagrammatic representation of the Green's
function for a process with a  $j$-pion intermediate
state.}    \label{fig-jpionint}  \end {figure}

\vfill \eject \null

\begin {figure} \vspace{27 mm} \hskip 30 mm \vskip 3 mm
\caption {The energy and momentum assignments in the exact
convolution integral result for one-pion exchange.}
\label{fig-FullCIOPE}  \end {figure}

\begin {figure} \vspace{45 mm} \hskip -10 mm \vskip 3 mm
\caption {The energy and momentum assignments in the
Kvinikhidze and Blankleider (KB) approximation to the exact
convolution integral expression for one-pion exchange, which
is shown in Figure \protect{\ref{fig-FullCIOPE}}.}
\label{fig-BKapOPE}    \end{figure}

\begin {figure} \vspace{25 mm} \hskip 45.5 mm \vskip 3 mm
\caption {An example of the diagrams left out of the
calculation in KB's approximation to one-pion exchange.}
\label{fig-BKOmit}  \end {figure}

\begin {figure} \vspace{25 mm} \hskip 45 mm \vskip 3 mm
\caption {An example of the diagrams included in the
calculation in KB's approximation to one-pion exchange.}
\label{fig-BKInc}  \end {figure}

\vfill \eject \null

\begin {figure} \vspace{25 mm} \hskip 48 mm \vskip 3 mm
\caption {A diagram containing the Jennings mechanism with
spectator pion which is included in KB's calculation of
one-pion exchange.}      \label{fig-BKInc2}   \end {figure}

\begin {figure} \vspace{35 mm} \hskip 25 mm \vskip 3 mm
\caption {The process on the left is always included in KB's
approximation to the exact convolution integral one-pion
exchange formula, while the process on the right is not always
included; it is forbidden in certain diagrams, e.g. Figure
\protect{\ref{fig-BKOmit}}, and allowed in others, e.g. Figure
\protect{\ref{fig-BKInc2}}.}       \label{fig-BKJenn}    \end
{figure}

\begin {figure} \vspace{70 mm} \hskip 50.5 mm \vskip 3 mm
\caption {These diagrams are an example of those omitted from
the calculation of two-pion exchange in the KLMK method. Note
that they are merely a different time-order of other included
diagrams, see e.g. Figure \protect{\ref{fig-KKinc}}.}
\label{fig-KKomit}     \end {figure}

\vfill \eject \null

\begin {figure} \vspace{68 mm} \hskip 13 mm \vskip 3 mm
\caption {These diagrams are included in the calculation of
two-pion exchange in the KLMK method.}
\label{fig-KKinc}    \end {figure}

\begin {figure} \vspace{40 mm} \hskip 3.5 mm \vskip 3 mm
\caption {The energy assignments in the KLMK method
calculation of two-pion exchange.}          \label{fig-TPE}
\end {figure}

\end {document}